\documentclass[10pt]{elsarticle}

\usepackage{amsthm,amsmath,amsfonts,amssymb,amscd,mathrsfs}
\usepackage{txfonts}
\usepackage{supertabular,soul}
\usepackage[usenames,dvipsnames]{xcolor}
\usepackage{tikz, graphicx,color,geometry}
 \usepackage{multirow}
 \usetikzlibrary{arrows}
\usepackage[pdftex,
            pdfauthor={Barrett},
            pdftitle={Automorphisms, Equitable Partitions, and Spectral Graph Theory},
            pdfsubject={Equitable Partitions, Spectral Graph Theory},
            pdfkeywords={Equitable Partitions, Spectral Graph Theory}]{hyperref}

\usepackage{bbm} 
\usepackage{hyperref}
\usepackage{yfonts}
\usepackage{eucal}
\usepackage{overpic}
\usetikzlibrary{calc}
\usepackage{enumitem}

\newcommand{\comment}[1]{}

\makeatletter
\def\ps@pprintTitle{%
\let\@oddhead\@empty
\let\@evenhead\@empty
\let\@oddfoot\@empty
\let\@evenfoot\@oddfoot}
\makeatother

\newtheorem{theorem}{Theorem}
\newtheorem{definition}{Definition}
\newtheorem{thm}{Theorem}[section]
\newtheorem{corollary}[thm]{Corollary}

\newtheorem{proposition}[thm]{Proposition}

\newtheorem{example}[thm]{Example}

\theoremstyle{remark}

\providecommand*{\propertyautorefname}{Property}

\let\oldmarginpar\marginpar
\renewcommand\marginpar[1]{\oldmarginpar[\raggedleft\footnotesize #1]%
{\raggedright\footnotesize #1}}

\begin{document}
\begin{frontmatter}

\date{\today}

\title{Link Prediction in Networks Using Effective Transitions}

\author[bryn]{Bryn Balls-Barker}
\address[bryn]{Department of Mathematics, Brigham Young University, Provo, UT 84602, USA, bryn.balls.barker@gmail.com}
\author[ben]{Benjamin Webb}
\address[ben]{Department of Mathematics, Brigham Young University, Provo, UT 84602, USA, bwebb@mathematics.byu.edu}

\begin{abstract}
We introduce a new method for predicting the formation of links in real-world networks, which we refer to as the method of effective transitions. This method relies on the theory of isospectral matrix reductions to compute the probability of eventually transitioning from one vertex to another in a (biased) random walk on the network. Unlike the large majority of link prediction techniques, this method can be used to predict links in networks that are directed or undirected which are either weighted or unweighted. We apply this method to a number of social, technological, and natural networks and show that it is competitive with other link predictors often outperforming them. We also provide a method of approximating our effective transition method and show that aside from having much lower temporal complexity, this approximation often provides more accurate predictions than the original effective transition method. We also, prove a number of mathematical results regarding our effective transition algorithm and its approximation.
\end{abstract}


\end{frontmatter}

\section{Introduction}
The study of networks has become increasingly relevant in the technological, natural, and social sciences. This is owing to the fact that many important systems can be described in terms of networks. In the social sciences, network theory facilitates the study of disease transmission \cite{Bog13},  the spread of information on the internet \cite{Joh04}, social media interactions \cite{Cha18}, etc. In the biological sciences network theory is used to understand properties and features of systems, including protein-protein interactions, metabolic modeling, and signal transmissions between macromolecules \cite{Pav11}. Network analysis is also used to study the interplay of the structure and function of technological systems such as the internet, power grids, and transportation networks \cite{Newman10}.

An essential feature of the large majority of these networks is that they have a dynamic \emph{topology}, that is a structure of interactions that evolves over time \cite{Bara00}. The structure of social networks, for instance, change as relationships are formed and dissolved. In information networks such as the WWW the network's topology changes over time as information is uploaded, linked, and updated. The topology of biological systems such as gene regulatory networks evolve as the network specializes the activity of existing genes to create new patterns of gene behavior \cite{Esp10}.

Although understanding the mechanisms that govern this structural evolution is fundamental to network science, these mechanisms, in general, are not well understood. Consequently, accurately predicting a network's eventual structure, function, or whether the network will fail at some future point in time are all largely out of reach for most real-world networks.

In an attempt to determine the mechanisms responsible for the changing structure of a network we are lead to the following link prediction problem: Given a network, which of the network's \emph{links}, i.e. interactions between existing network elements, are likely to form in the near future (see, for instance, \cite{link_prediction_social_networks}). Importantly, the link prediction problem can be used to study more than just which edges will appear in a network. It can also be used to predict which of the non-network edges are, in fact, in the network but currently undetected. Similarly, it can be used to detect which of the current network edges have been falsely assumed or determined to be a part of the network. More generally, any link prediction method can be used for community detection within a network \cite{Ghasemian18}.

Solutions to the link prediction problem have a number of applications. In national defense, link prediction is used in military organization to improve the decision making in battlefield situations and the design of operation plans \cite{Fan17}. In biology, link prediction is used in supervised prediction of gene networks \cite{Sho13} and is used to determine whether biochemical reactions are caused by specific sets of enzymes to infer causality \cite{Baz13}. Link prediction is also used to predict and control outbreaks of infectious diseases \cite{Mas13}. It is also of central importance to companies such as Facebook, Twitter and Google which need to know the current state and efficiently predict the future structure of the networks they use to accurately sort and organize data \cite{Que12}.

The barrier in determining whether network links truly exist in these and other settings is that testing and discovering interactions in a network requires significant experimental effort in the laboratory or in the field \cite{Cla08}. Similarly, determining experimentally when and where a new link will form may also be impractical, especially if the precise mechanism for link formation is unknown. For these reasons it is important to develop models for predicting the formation of links.

At present, there is an ever increasing number of proposed methods for predicting network links (see, for instance, \cite{Srin16}). Each method computes a \emph{score} for each possible link, i.e. network edge, or equivalently for each possible pair of nodes in the network. The links with the highest scores are those predicted to form. Not surprisingly, certain methods more accurately predict the formation of links in certain networks than in others. In practice, the majority of these link prediction methods can only be used on undirected networks, while a few others are more versatile and can accommodate directed networks as well as undirected networks. In fact, very few if any of these predictors can incorporate the edge weights of a network.

Here we introduce a new method that can be used to predict links in nearly any kind of network, including those that are directed or undirected and weighted or unweighted. For our method the score function is based on the notion of an \emph{effective transition}, which is the probability of eventually transitioning from node $i$ to node $j$ before returning the node $i$. Specifically, our method takes a transition matrix associated with the network and from it creates an \emph{effective transition matrix}, which we use to predict which links will form next in the network.

We show that our method of effective transitions is competitive with other commonly used predictors on a wide variety of networks often outperforming them. Because of the method's relatively high temporal complexity we also derive an approximation of this method which has a much lower computational cost. We similarly demonstrate that this approximation is competitive with other link predictors and in many cases outperform the original version of this method.

Additionally, we prove a number of results regarding the effective transition method and its approximations. This includes proving that the effective transition matrix can be computed using isoradial reductions, which are matrix transformations that preserve the spectral radius and other properties of a matrix (see Theorem \ref{thm:eql}). We also show that the effective transition matrix inherits a number of properties from the original transition matrix, e.g. nonnegativity, irreducibility, the same unique stationary distribution in the case of stochastic matrices (see Proposition \ref{prop:stoch}). Additionally, we prove that our approximation method converges monotonically to our original effective transition method (see Theorem \ref{thm:lstep}).

The article is structured as follows. In Section \ref{sec:graph} we introduce the notation and mathematical concepts used to describe networks. In Section \ref{sec:preds} we give an overview of common link predictors. In Section \ref{sec:net} we formally introduce the method of effective transitions and describe how these transitions can be computed using isoradial reductions. In Section \ref{sec:strong} we apply our effective transition methods to a number of real-world networks and compare our predictions to those of the standard link predictors described in Section \ref{sec:preds}. In Section \ref{sec:approx} we give an approximation method for computing effective transition scores and in Section \ref{sec:weight} we use our effective transition methods to predict links on weighted networks. Section \ref{sec:conc} has some concluding remarks which is followed by an Appendix containing the proofs of the main results of the paper.

\section{Preliminaries}\label{sec:graph}

The standard method used to describe the topology of a network is a graph. Here, a \emph{graph} $G=(N,E)$ is composed of a \emph{node set} $N$ and an \emph{edge set} $E$. The node set $N$ represents the \emph{elements} of the network, while the edges $E$ represent the links or \emph{interactions} between these network elements. In social networks, for example, the nodes are typically individuals where edges represent specific types of relationships between these individuals, e.g. friendships between Facebook users. In the World Wide Web, which is an example of an information network, the nodes represent webpages and the edges represent hyperlinks between these pages. In neural networks, which are the network of neurons in the brain, nodes represent neurons and edges represent physical connections formed by synapse between the neurons.

In some networks there is a direction to each interaction. For instance, in a citation network, in which network elements are papers and edges represent whether one paper cites another, a paper can only cite existing papers. Thus, each edge is \emph{directed} having a clearly defined direction. A network in which interactions are directed is referred to as a \emph{directed network}. If the network's interactions are \emph{undirected}, the network is an \emph{undirected network}. For example, Facebook is an undirected network in which friendships are mutual and are, therefore, represented by undirected edges.

Throughout this paper, we let $e_{ij} \in E$ denote the directed edge from node $i$ to node $j$ in a directed graph $G = (N, E)$. If $G$ is undirected, then $e_{ij} = e_{ji}$ denotes the undirected edge between node $i$ and node $j$. If $G=(N,E)$ is undirected then $G$ is \emph{connected} if for all $i, j \in N$ with $i \neq j$, there is a path in $G$ from node $i$ to node $j$. Otherwise, $G$ is \emph{disconnected}. If $G$ is directed, we call $G$ \emph{strongly connected} if for all $i,j \in N$ with $i \neq j$ there exists a directed path from node $i$ to node $j$. The directed graph $G$ is \emph{weakly connected} if, when we undirect each of its edges, the resulting graph is connected. If a directed graph $G$ is not weakly connected, it is \emph{disconnected}.

In the following section we describe a number of the most common link predictors used to identify and predict links in networks. We split these predictors into two categories depending on whether they can be used on (i) both directed and undirected networks or (ii) only on undirected networks.

\section{Link Predictors}\label{sec:preds}

As previously mentioned, there are many existing link predictors. Here we describe some of the most common of these. Later, we will compare these against the \emph{effective transition predictors} we introduce in this paper. These standard predictors are divided into two categories based on what type of networks they can be applied to (see Sections \ref{sec:directed} and \ref{sec:undirected}). For a network given by the graph $G=(N,E)$ we define a score function $score(i,j)$ to be the score assigned by the predictor to the edge $e_{ij} \in E$ or to the \emph{potential edge} $e_{ij} \notin E$ for all $i, j \in N$.

When predicting which links will form in a given network, the potential edges with the $\kappa$ highest scores are predicted to form for some predetermined integer $\kappa>0$. If $e_{ij},e_{k\ell} \notin E$ and $score(i,j)>score(k,\ell)$ then we write $e_{ij} \succ e_{k\ell}$. We can similarly use $score(i,j)$ to predict which of the network edges are not part of the network by selecting those edges of the network with the $\kappa>0$ lowest scores, although our main focus in this paper will be predicting which of the potential edges will form.

\subsection{Link Prediction for Directed or Undirected Networks}\label{sec:directed}
The following predictors can be used on both directed and undirected networks (see \cite{link_prediction_social_networks} for more details). For each predictor, the associated graph $G=(N,E)$ is assumed to be connected if $G$ is undirected and strongly connected if $G$ is directed.

\textit{Shortest Path:} This predictor computes the length of the shortest path from node $i$ to node $j$ and sets the negative length as the score
$$score(i,j) = \textrm{negated length of shortest path from node i to node j}.$$

\textit{Katz:} The Katz metric counts all possible paths between two nodes and discounts the longer paths exponentially. Let $path^\ell_{ij}$ be the set of all paths of length $\ell$ from node $i$ to node $j$. Then, given a weight $0<\beta<1$, the Katz score is
$$score(i,j) = \sum\limits_{\ell=1}^\infty \beta^\ell |path^\ell_{ij}|.$$

\textit{Hitting Time:} The hitting time from node $i$ to node $j$,  denoted $H_{ij}$, is the expected number of steps required to reach node $j$ in a random walk starting at node $i$. The related score is given by
$$score(i,j) = -H_{ij}.$$

\subsection{Link Prediction for Undirected Networks}\label{sec:undirected}
The following predictors can only be used on undirected networks. Similar to the graphs in Section \ref{sec:directed}, it is assumed that each graph is connected.

\textit{Common Neighbors:} In an undirected graph $G$, nodes $i$ and $j$ are \textit{adjacent} if $e_{ij} \in E$, or equivalently if $e_{ji} \in E$. The score using common neighbors is given by the number of adjacent nodes two nodes have in common:
$$score(i,j)=|\Gamma(i)\cap \Gamma(j)|,$$ where $\Gamma(i)$ is the set of all nodes adjacenct to node $i\in N$.

\textit{Jacaard's Coefficient:} Jacaard's coefficienct is a normalized version of \emph{common neighbors} that incorporates the total number of neighbors of both nodes given by $$score(i,j)=\frac{|\Gamma(i)\cap \Gamma(j)|}{|\Gamma(i)\cup \Gamma(j)|}.$$

\textit{Preferential Attachment:} Preferential attachment is based on the idea that highly connected nodes (nodes with the most neighbors) are more likely to form links, an observed pattern in many real-world 
networks \cite{New01}. This leads to the preferential attachment score $$score(i,j)=|\Gamma(i)|| \Gamma(j)|.$$

\textit{Resistance Distance:} Resistance Distance is a scaled, undirected version of hitting time. It has the following score function $$score(i,j)=L^\dagger_{ii}+L^\dagger_{jj}-2L^\dagger_{ij},$$
where $L^\dagger$ is the Moore-Penrose inverse of the Laplacian Matrix $L$ [ref Laplacian]. (For an approximation of this score function with low temporal complexity, see \cite{Pach17}.)

\section{Effective Transitions}\label{sec:net}

In this section we introduce a new method of link prediction which we refer to as \emph{effective transitions}. We begin with a graph $G=(N,E)$ with nodes $N=\{1,2,\dots,n\}$ and an associated \emph{transition matrix} $P=[p_{ij}] \in \mathbb{R}^{n\times n}$.
\begin{definition} \label{def:trans} \textbf{(Transition Matrix)}
A matrix $P=[p_{ij}] \in \mathbb{R}^{n\times n}$ associated with a graph $G=(N,E)$ is a \emph{transition matrix} if $p_{ij} > 0 $ whenever $e_{ij} \in E$ and where $p_{ji} = 0$ otherwise.
\end{definition}

For the sake of motivation, for the moment we assume that the transition matrix $P$ is row stochastic where $p_{ij}$ gives the probability of ``transitioning'' from node $i$ to node $j$.  In a social network $p_{ij}$ could represent the probability that information is passed from individual $i$ to individual $j$ (cf. Example \ref{ex:1}). In the World Wide Web $p_{ij}$ could be the probability of taking a hyperlink from webpage $i$ to webpage $j$. In a neuronal network $p_{ij}$ could be the probability that neuron $j$ fires immediately after neuron $i$.

\subsection{Motivation}
For a network given by the graph $G$ and an associated stochastic transition matrix $P=[p_{ij}]$, we can consider a random walk on $G$ in which a walker moves through the network by moving at each time step from the current node $i$ to the node $j$ with probability $p_{ij}$. That is, we have a Markov chain with states $N=\{1,2,\dots,n\}$ in which we transition from state $i$ to state $j$ with probability $p_{ij}$. Note that if the graph $G$ is (strongly) connected then the matrix $P$ is \emph{irreducible} and the walker can \emph{eventually transition} from any node to any other node in the network via a sequence of single-step transitions. With this notion of an eventual transition in mind, we can use the stochastic transition matrix $P$ to compute an associated \textit{effective transition matrix}, which is defined as follows.

\begin{definition}\label{def:etm} \textbf{(Stochastic Effective Transition Matrix)}
For a graph $G=(N,E)$, let $P=[p_{ij}]\in\mathbb{R}^{n\times n}$ be an associated row stochastic transition matrix. The \textit{effective transition matrix} $\mathcal{E}(P)=[\varepsilon_{ij}]\in\mathbb{R}^{n\times n}$ is the matrix in which $\varepsilon_{ij}$ is the probability of eventually transitioning from node $i$ to node $j$ before returning to node $i$ for $i \neq j$. The diagonal entries of the matrix $\mathcal{E}(P)$ are given by
\begin{equation}\label{efftrans}
\varepsilon_{ii}=\sum\limits_{k=1,k\neq i}^n(1-\varepsilon_{ik}) \ \ \text{for} \ \ i\in N,
\end{equation}
which is the sum of the probabilities $1-\varepsilon_{ik}$ of eventually returning to node $i$ before transitioning to node $k$, over all $k\neq i$.
\end{definition}

We use the effective transition matrix $\mathcal{E}(P)$ to create a score function, which is a special case of the \emph{effective transition predictor}. (See Section \ref{Sec:Gen} for the general definition of the effective transition predictor.)
\vskip .2cm
\emph{Stochastic Effective Transitions:} Suppose $G = (N,E)$ is a (strongly) connected graph with row stochastic transition matrix $P$. The effective transition score is
\begin{equation}\label{eqn:score1}
score(i,j) = \varepsilon_{ij}
\end{equation}
where $\mathcal{E}(P) = [\varepsilon_{ij}]$ is the effective transition matrix associated with $P$.

In real-world networks, edges going from a node to itself, called \emph{loops}, are rarely used. This is often the case because the meaning of such edges do not make sense within the context of the network. In such cases we modify the score given by Equation \ref{eqn:score1} so that \begin{equation}\label{eqn:score2}
score(i,j) = \begin{cases}
\varepsilon_{ij} &\text{if } i\neq j, \\
0 &\text{otherwise.}
\end{cases}
\end{equation}

It is worth mentioning that a transition matrix $P$ is associated with exactly one graph $G=(N,E)$, since $e_{ij} \in E $ if $p_{ij}>0$ and $p_{ij}=0$, otherwise. Hence, there is one graph whose transition matrix is $\mathcal{E}(P)$, which we denote by $\mathcal{E}(G)$ and refer to it as the \emph{effective graph (network)} of $G$ associated with $P$.

To illustrate the effective transition predictor we consider the following example.

\begin{figure}
\begin{center}
  \begin{overpic}[scale=.575]{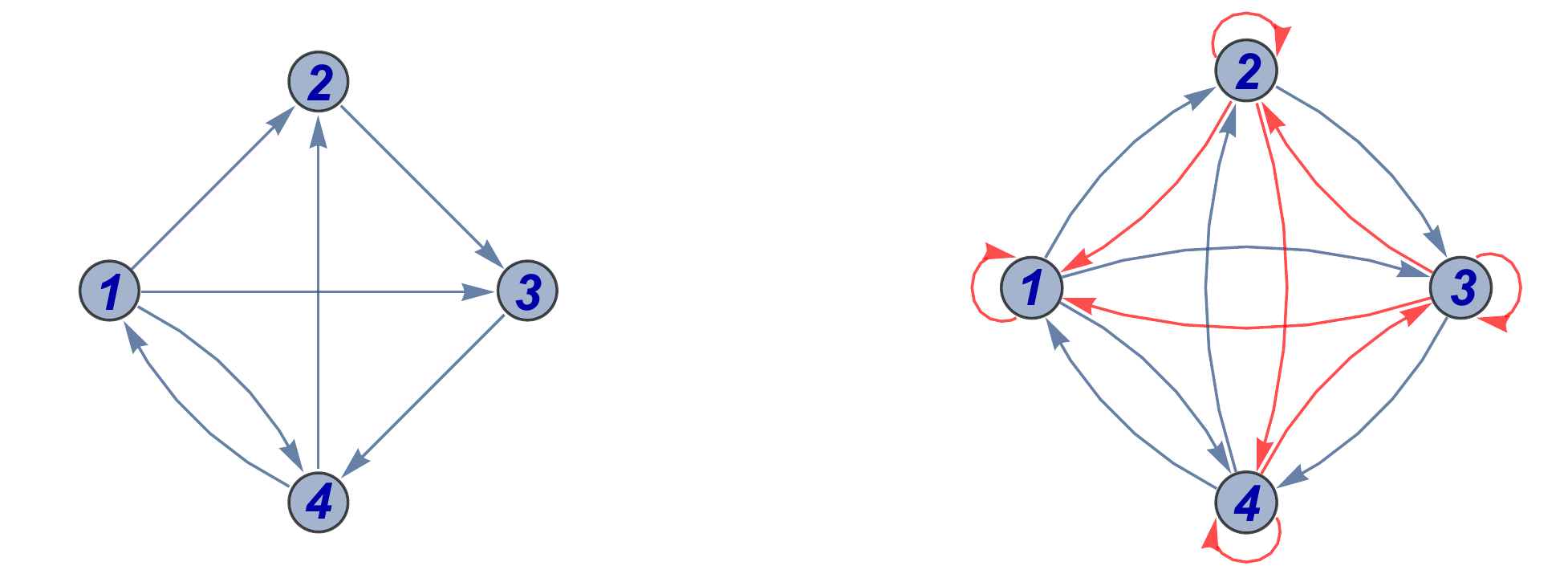}
    \put(19,-3){$G$}


    \put(76,-3){$\mathcal{E}(G)$}
    \end{overpic}
    \vspace{0.2cm}
  \caption{The network $G=(N,E)$ consisting of four individuals $N=\{1,2,3,4\}$ is shown (left), with transition matrix $P=[p_{ij}]\in\mathbb{R}^{4\times 4}$ given by Equation \eqref{eq:mat1}. Here $p_{ij}$ indicate that probabilities that information is passed directly from one member of the network to another. The associated effective network $\mathcal{E}(G)$ is shown (right) which has the effective transition matrix $\mathcal{E}(P)=[\varepsilon_{ij}]\in\mathbb{R}^{4\times4}$ given by Equation \eqref{eq:mat2}. The probabilities $\varepsilon_{ij}$ indicate the probability that information is eventually passed from one individual to another in $G$. In $\mathcal{E}(G)$ red edges indicate the new edges, e.g. potential edges,  that do not belong to the original network.}\label{Fig:0}
\end{center}
\end{figure}

\begin{example}\label{ex:1} \textbf{(Information Transfer in Social Networks)}
Consider the social network consisting of the four individuals $N=\{1,2,3,4\}$, represented by the graph $G=(N,E)$ in Figure \ref{Fig:0} (left) with transition matrix $P=[p_{ij}]\in\mathbb{R}^{4\times 4}$ given by
\begin{equation}\label{eq:mat1}
P=
\left[
\begin{array}{cccc}
0&\frac{1}{3} & \frac{1}{3} & \frac{1}{3}\\ [1pt]
0 & 0 & 1 & 0\\ [1pt]
0 & 0 & 0 & 1\\ [1pt]
\frac{1}{2} & \frac{1}{2} & 0 & 0
\end{array}
\right].
\end{equation}
Here $p_{ij}$ represents the probability that individual $i$ passes information directly to individual $j$. For this network the effective transition matrix $\mathcal{E}(P)$ is given by
\begin{equation}\label{eq:mat2}
\mathcal{E}(P)=
\left[
\begin{array}{cccc}
\frac{1}{2} &\frac{2}{3} & \frac{5}{6} & 1\\ [1pt]
\frac{1}{2} & \frac{1}{2} & 1 & 1\\ [1pt]
\frac{1}{2} & \frac{4}{5} & \frac{7}{10} & 1\\ [1pt]
\frac{1}{2} & \frac{2}{3} & \frac{5}{6} & 1
\end{array}
\right],
\end{equation}
where the nondiagonal entries of $\mathcal{E}(P)$ represent the probabilities that information is eventually passed from the individual $i$ to another individual $j$ before this information returns to individual $i$. The effective graph $\mathcal{E}(G)$ of this network is shown in Figure \ref{Fig:0} (right), in which red edges indicate edges not present in the original network.

Using the effective transition score defined in Equation \ref{eqn:score2} we have the ranking
\begin{equation}\label{eq:1}
e_{24} \succ e_{43} \succ e_{32}  \succ e_{21},  e_{31}
\end{equation}
of the potential nonloop edges in $G$ and the ranking
\begin{equation}\label{eq:2}
e_{14},e_{23},e_{34} \succ e_{13} \succ e_{12},e_{42} \succ e_{41}
\end{equation}
of those edges in $G$. Here we choose to ignore loops as the notion of passing information from an individual back to themselves does not make sense in this example. The ranking given in \eqref{eq:1} can be used to predict which edge(s) are most likely to form in $G$, which would be new social links through which information could be directly passed. The ranking given by \eqref{eq:2} can similarly be used to predict which edge(s) in $G$ have been falsely identified as being part of the network. (Here, the edge $e_{41}$ is the edge predicted to be the least likely edge of the network by \eqref{eq:2}.)
\end{example}

\subsection{Computation of Effective Transition Matrices}
In Example \ref{ex:1} we had a very small network from which we could, by hand, compute an effective transition matrix. Most real networks, however, are much larger consisting of hundreds, thousands, or more nodes. To compute the effective transition scores for such networks we require an algorithm that can scale to the size of these real-world networks.

To compute the effective transition matrix $\mathcal{E}(P)$ associated with a large matrix $P\in\mathbb{R}^{n\times n}$, we use a matrix transform referred to as an isoradial matrix reduction, or the Perron compliment \cite{Mey89}. An isoradial reduction is a special type of isospectral reduction \cite{BWBook} that preserves a number of spectral properties of a matrix, including its spectral radius. See \cite{Smith19} for more details.

To define an isoradial matrix reduction we require the following. For a matrix $M\in\mathbb{R}^{n\times n}$ let $N=\{1,\ldots,n\}$. If the sets $R,C\subset N$ are proper subsets of $N$, we denote by $M_{RC}$ the $|R| \times |C|$ \emph{submatrix} of $M$ with rows indexed by $R$ and columns indexed by $C$. We denote the subset of $N$ not contained in $S$ by $\bar{S}$, that is $\bar{S}$ is the \emph{complement} of $S$ in $N$. Using this notation, an isoradial reduction of a square real-valued matrix is defined as follows.

\begin{definition} \textbf{(Isoradial Reductions)}
The \emph{isoradial reduction} of a matrix $M\in\mathbb{R}^{n\times n}$ over a nonempty subset $S\subset N$ is the matrix
\begin{equation}\label{eqn:IRR}
\mathcal{I}_{S}(M)=M_{SS}-M_{S\bar{S}}\left(M_{\bar{S}\bar{S}}-\rho(M) I\right)^{-1}M_{\bar{S}S}\in\mathbb{R}^{|S|\times|S|},
\end{equation}
where $\rho(M) = \max \{|\lambda |: \lambda \textrm{ is an eigenvalue of } M\}$ is the \emph{spectral radius} of $M$.
\end{definition}

The isoradial reduction $\mathcal{I}_S(M)$ does not exist for every square real-valued matrix $M$ and every subset $S \subseteq N$, due to the fact that the inverse taken in Equation \eqref{eqn:IRR} may not exist. However, a nonnegative irreducible matrix will always have an isoradial reduction over any subset $S\subseteq N$ (see Theorem \ref{thm:irr} in the Appendix). This allows us to compute the effective transition matrix for a network as follows.

\begin{theorem}\label{thm:eql} \textbf{(Computing Stochastic Effective Transitions)}
Suppose the (strongly) connected graph $G=(N,E)$ has the row stochastic transition matrix $P\in\mathbb{R}^{n\times n}$. Then the effective transition matrix $\mathcal{E}(P)=[\varepsilon_{ji}]\in\mathbb{R}^{n\times n}$ is given by
\begin{equation}
\label{eqn:stoch}
\varepsilon_{ij}=
\begin{cases}
    \mathcal{I}_{\{i,j\}}(P)_{12} &\text{if} \ i< j\\
    \mathcal{I}_{\{i,j\}}(P)_{21} &\text{if} \ i> j\\
\sum\limits_{k>i} \mathcal{I}_{\{i,k\}}(P)_{11}+\sum\limits_{k<i} \mathcal{I}_{\{i,k\}}(P)_{22} &\text{if} \ i=j
\end{cases}
\end{equation}
where $\mathcal{I}_{\{i,k\}}(P)\in \mathbb{R}^{2\times 2}$ is the isoradial reduction of $P$ over the pair $\{i,k\}\subseteq N$.
\end{theorem}

As Theorem \ref{thm:eql} relies on the theory of isospectral reductions its proof is given in the Appendix, where the necessary details of this theory are found (see also \cite{BWBook}).

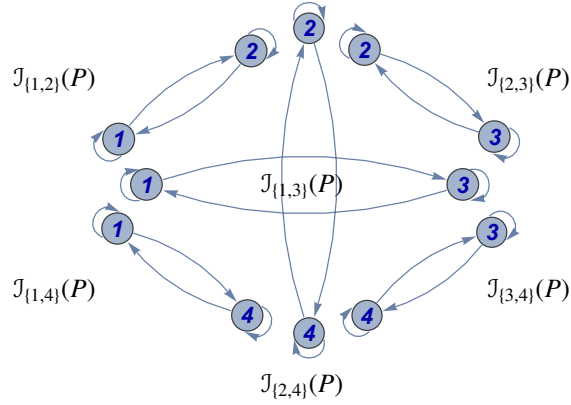
\begin{figure}
\begin{center}
  \begin{overpic}[scale=.475]{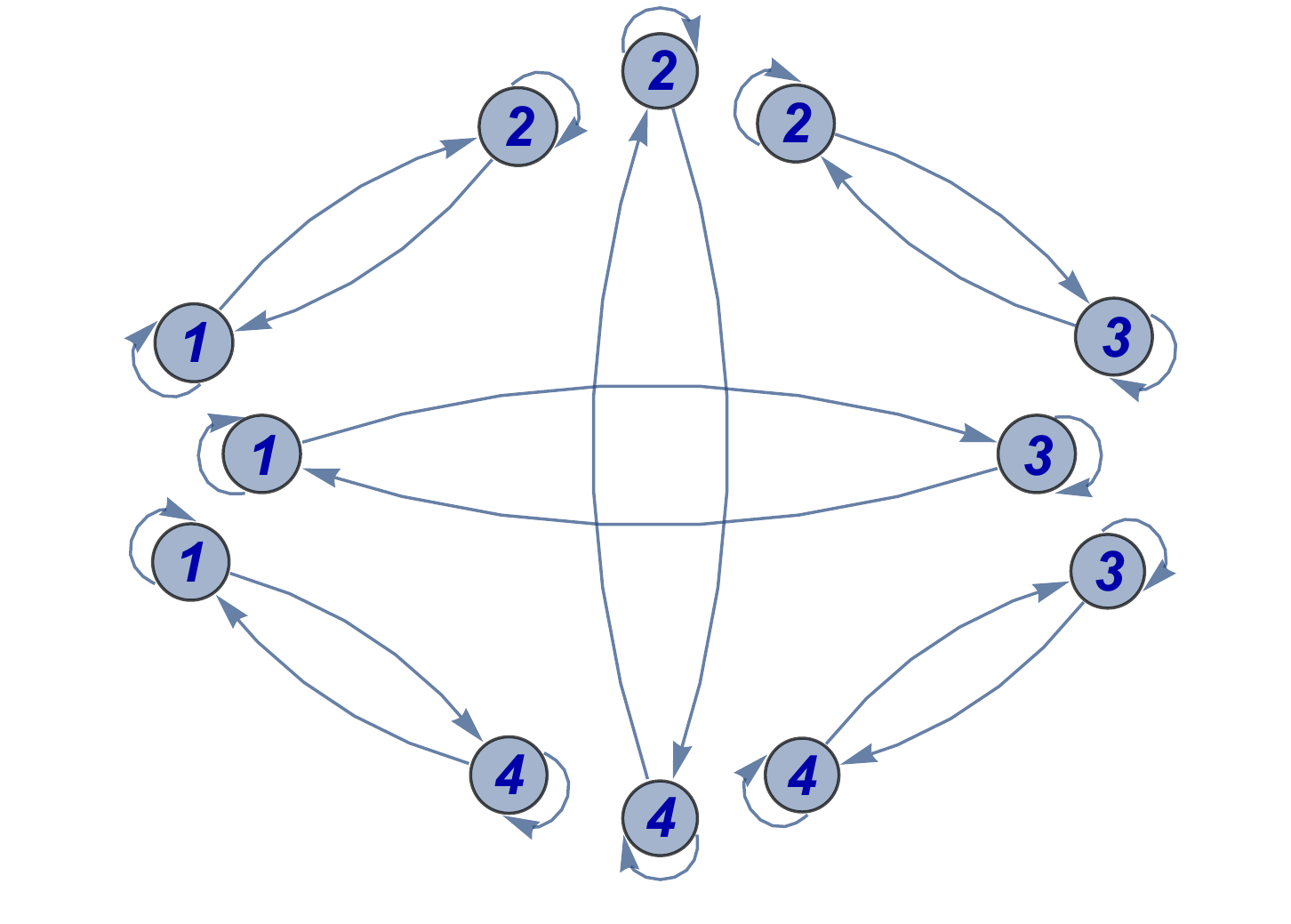}

  \put(-5,55){$\mathcal{I}_{\{1,2\}}(P)$}
  \put(42,35){$\mathcal{I}_{\{1,3\}}(P)$}
  \put(-5,15){$\mathcal{I}_{\{1,4\}}(P)$}

  \put(85,55){$\mathcal{I}_{\{2,3\}}(P)$}
  \put(42,-3){$\mathcal{I}_{\{2,4\}}(P)$}
  \put(85,15){$\mathcal{I}_{\{3,4\}}(P)$}



    \end{overpic}
    \end{center}
  \caption{The graphs representing the six isoradial reductions $\mathcal{I}_{\{i,j\}}(P)$ for $1\leq i<j\leq 4$ of the network $G$ in Figure \ref{Fig:0} (left) are shown with individual transition matrices given in Example \ref{ex:comp}. Merging each of the copies of node $i$ and adding the weights of their respective loops for $i=1,2,3,4$ yields the effective network $\mathcal{E}(G)$ shown in Figure \ref{Fig:0} (right).}\label{Fig:1}
\end{figure}

\begin{example}\label{ex:comp}
Consider the social network given by the graph $G$ in Figure \ref{Fig:0} (left) with transition matrix $P$ given in Equation \ref{eq:mat1}. To compute the effective transition matrix $\mathcal{E}(P)$ using Theorem \ref{thm:eql} we first compute each of the six isoradial reductions $\mathcal{I}_{\{i,j\}}(P)$ for $1\leq i<j \leq 4$. The results are the six $2 \times 2$ matrices
\[
\mathcal{I}_{12}(P)=
\left[
\begin{array}{cc}
\frac{1}{3} & \frac{2}{3}\\ [1pt]
\frac{1}{2} & \frac{1}{2}
\end{array}
\right], \
\mathcal{I}_{13}(P)=
\left[
\begin{array}{cc}
\frac{1}{6} & \frac{5}{6}\\ [1pt]
\frac{1}{2} & \frac{1}{2}
\end{array}
\right], \
\mathcal{I}_{14}(P)=
\left[
\begin{array}{cc}
0 & 1\\ [1pt]
\frac{1}{2} & \frac{1}{2}
\end{array}
\right]
\]
\[
\mathcal{I}_{23}(P)=
\left[
\begin{array}{cc}
0 & 1\\ [1pt]
\frac{4}{5} & \frac{1}{5}
\end{array}
\right], \
\mathcal{I}_{24}(P)=
\left[
\begin{array}{cc}
0 & 1\\ [1pt]
\frac{2}{3} & \frac{1}{3}
\end{array}
\right], \
\mathcal{I}_{34}(P)=
\left[
\begin{array}{cc}
0 & 1\\ [1pt]
\frac{5}{6} & \frac{1}{6}
\end{array}
\right]
\]
corresponding to the six two-vertex graphs shown in Figure \ref{Fig:1}. Merging these graphs into a single graph by identifying the three copies of the node $i$ for $i=1,2,3,4$ and adding the weights of their respective loops gives us the graph $\mathcal{E}(G)$ shown in Figure \ref{Fig:0} (right) or equivalently the matrix $\mathcal{E}(P)$ given in Equation \eqref{eq:mat2}.
\end{example}

The effective transition matrices that arise from stochastic transition matrices have properties related to random walks on a network or, equivalently, to the associated Markov chain. To describe these properties suppose $P \in \mathbb{R}^{n \times n}$ is an irreducible row stochastic matrix associated with the graph $G$. Then the matrix
\[
\mathcal{S}(P)=\frac{1}{n-1}\mathcal{E}(P)\in\mathbb{R}^{n\times n},
\]
which we refer to as the \emph{scaled} effective transition matrix of $P$, is also a row stochastic matrix. The reason is that each entry of $\mathcal{E}(P)=[\varepsilon_{ij}]$ is nonnegative and each row of $\mathcal{E}(P)$ sums to $(n-1)$ via Equation \eqref{efftrans}. Using this we let $\mathcal{S}(G)$ be the effective network associated with the effective transition matrix $\mathcal{S}(P)$.

Since $\mathcal{S}(P)$ is a scaled version of $\mathcal{E}(P)$ then both matrices induce the same ranking on the edges of the associated network. Therefore, the effective transition ranking on $G$ induced by the stochastic transition matrix $P$ can be given in terms of another stochastic matrix, namely $\mathcal{S}(P)$. Moreover,  note that if $P\in\mathbb{R}^{n\times n}$ is a row stochastic transition then a vector $\pi\in\mathbb{R}^n$ is a \emph{stationary distribution} of the associated Markov chain if $\pi P=\pi$ and $\sum_{i=1}^n\pi_i=1$. If $P$ is irreducible then $\pi$ is unique. If $P$ is \emph{primitive}, that is some power of $P$ is positive, then $P$ is irreducible and any initial distribution converges to the Markov chain's unique stationary distribution. In this case the component $\pi_i$ of the stationary distribution gives the probability that a random walker will be at node $i$ on a long walk on the associated graph $G$.

A fact we will later prove is that if $P$ is irreducible then both $P$ and $\mathcal{S}(P)$ have the same unique stationary distribution and $\mathcal{S}(P)$ is primitive (see Proposition \ref{prop:stoch} (iii)). Hence, if $P$ is primitive then on a random walk through either $G$ or $\mathcal{S}(G)$ the probability of being at node $i$ in either graph is asymptotically the same irrespective of the walkers initial probability distribution.

\subsection{Generalizing Effective Transition}\label{Sec:Gen}
As previously mentioned, the transition matrix associated with a given network need not be row stochastic. In fact, as described in \ref{def:trans}, a matrix $M=[m_{ij}]$ is a \emph{transition matrix} associated with a network $G=(N,E)$ if $m_{ij}>0$ when $e_{ij} \in E$ and $m_{ij}=0$ otherwise. That is, $M$ need not be stochastic. By analogy to Equation \eqref{eqn:stoch}, we define the effective transition matrix $\mathcal{E}(M)$ associated with a general transition matrix $M$ as follows.

\begin{definition} \label{def:etm2} \textbf{(General Effective Transition Matrix)}
For a (strongly) connected graph $G=(N,E)$, let $M\in\mathbb{R}^{n\times n}$ be an associated transition matrix. The \emph{effective transition matrix} $\mathcal{E}(M)=[\varepsilon_{ij}]\in\mathbb{R}^{n\times n}$ associated with $M$ is given by
\begin{equation}\label{eqn:def}
\varepsilon_{ij}=
\begin{cases}
    \mathcal{I}_{\{i,j\}}(M)_{12} &\text{if} \ i< j\\
    \mathcal{I}_{\{i,j\}}(M)_{21} &\text{if} \ i> j\\
\sum\limits_{k>i} \mathcal{I}_{\{i,k\}}(M)_{11}+\sum\limits_{k<i} \mathcal{I}_{\{i,k\}}(M)_{22} &\text{if} \ i=j
\end{cases}
\end{equation}
where $\mathcal{I}_{\{i,k\}}(M)\in \mathbb{R}^{2\times 2}$ is the isoradial reduction of $M$ over the pair $\{i,k\}\subseteq N$.
\end{definition}

Definition \ref{def:etm2} allows us to define a general effective transition predictor for any network with an associated transition matrix $M$, whether or not this matrix is stochastic.

The reason we introduced the notion of an effective transition matrix first in terms of a stochastic matrix was to motivate this more general notion of an effective transition matrix. However, we will see in Section \ref{sec:strong} that it depends on the particular network whether a stochastic or non-stochastic transition matrix is more effective in predicting which links will form next in a network.

To describe a number of properties of effective transition matrices we note that if the transition matrix $M\in\mathbb{R}^{n\times n}$ is irreducible then the Perron-Frobenius theorem implies that its spectral radius $\rho(M)$ is an algebraically simple eigenvalue. Hence, there is a unique positive eigenvector $\mathbf{v}\in\mathbb{R}^n$ such that $M\mathbf{v}=\rho(M)\mathbf{v}$ and $\sum_{i=1}^n v_i=1$. In this case we refer to this vector as the \emph{leading eigenvector} of $M$.

\begin{proposition}\label{prop:stoch} \textbf{(Properties of Effective Transition Matrices)}
Suppose $M\in\mathbb{R}^{n \times n}$ is an irreducible nonnegative matrix. Then\\
(i) $\mathcal{E}(M)$ is nonnegative and irreducible;\\
(ii) $M$ and $\mathcal{E}(M)$ have the same leading eigenvector; and\\
(iii) the spectral radius $\rho(\mathcal{E}(M))=(n-1)\rho(M)$.\\
If $M$ is a row stochastic matrix then\\
(iv) the Markov chains associated with $M$ and $\mathcal{S}(M)$ have the same unique stationary distribution and $\mathcal{S}(M)$ is primitive.
\end{proposition}

Properties (i) and (ii) of Proposition \ref{prop:stoch} state that the effective transition matrix $\mathcal{E}(M)$ inherits nonnegativity, irreducibility, and its leading eigenvector from $M$. However, the spectral radius of the two is not the same for $n\neq2$. An exception to this is the case in which $M$ is row stochastic, in which case $\rho(M)=\rho(\mathcal{S}(M))=1$ as $\mathcal{S}(M)$ is row stochastic. Here, property (iii) states that beyond this the Markov chains associated with $M$ and $\mathcal{S}(M)$ have the same unique stationary distribution and $\mathcal{S}(M)$ is primitive even if $M$ is not.

\subsection{Deriving Transition Matrices}\label{sec:ET}

In most real-world networks, we are not specifically provided with a transition matrix. However, there are a few natural candidates that we can use that rely on the adjacency matrix and degree matrix of the network. Given a network with graph $G=(N,E)$, the \emph{adjacency matrix} $A= [a_{ij}] \in \{0,1\}^{n\times n}$ of $G$ is the matrix with entries given by
\[
a_{ij}=
\begin{cases}
1  &\text{if} \ e_{ij} \in E\\
0 &\text{otherwise.}
\end{cases}
\]
The \emph{degree matrix} $D$ of the graph $G$ is the diagonal matrix $D = diag(d_1, d_2, ... , d_n)$ in which $d_i = \sum_{j=1}^n a_{ij}$ is the degree of node $i$, or equivalently the number of outgoing edges from node $i$.

For a (strongly) connected graph both $A$ and $AD^{-1}$ are natural choices for the transition matrix of the graph. When $A$ is used we call this our \emph{standard effective transition method}. When $AD^{-1}$ is used, we call this the \emph{normalized effective transition method}. We refer to this second method as ``normalized'' because $AD^{-1}$ is the matrix $A$ in which the $i$th row is divided by the degree $d_i$. The result is a row stochastic matrix in which the nonzero entries in a column are the same. (Note this is the matrix used in computing the PageRank centrality of a node, which is the basis of the algorithm employed by Google \cite{Newman10}.)

\begin{example}
If the adjacency matrix $A\in\mathbb{R}^{4\times 4}$ of the network given by $G$ in Figure \ref{Fig:0} (left) is used as the network's transition matrix then $A$ and the associated effective transition matrix are
\[
A=
\left[
\begin{array}{cccc}
0 & 1 & 1 & 1\\
0 & 0 & 1& 0\\
0 & 0 & 0 & 1\\
1 & 1 & 0 & 0
\end{array}
\right] \ \ \text{and} \ \
\mathcal{E}(A)=
\left[
\begin{array}{cccc}
\frac{1+\sqrt{5}}{2} & 2 & 2 & 2\\
\frac{3-\sqrt{5}}{2} & \frac{3-\sqrt{5}}{2} & 1 & \frac{-1+\sqrt{5}}{2}\\
\frac{-1+\sqrt{5}}{2} & \frac{1+\sqrt{5}}{2} & 1 & 1\\
1 & \frac{1+\sqrt{5}}{2} & \frac{1+\sqrt{5}}{2} & 2
\end{array}
\right],
\]
respectively. Here the effective transition matrix $\mathcal{E}(A)$ induces the ranking
\begin{equation}\label{eq:3}
e_{32},e_{43}  \succ e_{24},e_{31} \succ e_{21}
\end{equation}
of the non-loop edges not in $G$ and the ranking
\begin{equation}\label{eq:4}
e_{12},e_{13},e_{14} \succ e_{42} \succ e_{23},e_{34},e_{41}
\end{equation}
of the non-loop edges in $G$. This is the ranking given by the standard effective transition method. The normalized effective transition method is, in fact, the one carried out in Example \ref{ex:1} as $AD^{-1}$ is the stochastic matrix $P$ given in Equation \eqref{eq:mat1}. We note that the rankings given by the standard and normalized effective transition methods differ in a number of ways and it is an open question as to what the similarities and differences are between such rankings (cf. Equations \eqref{eq:1} and \eqref{eq:3}). Also, one can check that $\mathcal{E}(A)$ is nonnegative, irreducible, has the same leading eigenvector as $A$, and $\rho(\mathcal{E(A)})=3\rho(A)=3(1+\sqrt{5})/2$ as guaranteed by Proposition \ref{prop:stoch}.
\end{example}



\section{Predictions and Accuracy of Effective Transitions on Real Networks}\label{sec:strong}

In this section we apply both our standard and normalized effective transition method to a number of real-world networks and compare accuracy of these predictions against the other link predictors described in Section \ref{sec:preds}. In each case the networks we select are either connected or strongly connected depending on whether they are undirected or directed, respectively.

\subsection{Methods and Predictions}
To test and compare the effective transition method against other link predictors we start with a network given by a graph $G=(N,E)$, then carry out the following steps:
\vspace{.2cm}

\noindent\emph{Step 1:} We begin by first splitting the network's edges into training and test sets. The training set contains the first 80\% of edges formed in the network and the test set contains the remaining 20\% of the edges. We let $\kappa$ be the number of edges in the test set.\\

\noindent\emph{Step 2:} We remove any edges from the test set that involve nodes not included in the test set. (This is because our algorithm, and all other algorithms we are comparing, are unable to predict edges involving nodes not included in the training set.)\\

\noindent\emph{Step 3:} We use each applicable link predictor from Section \ref{sec:preds} to predict the next $\kappa$ edges that form within the network, or equivalently the 20\% of the network edges contained in the test set. The accuracy of each predictor is given by the percentage of correct predictions, i.e. those predicted edges that did in fact form.

\subsection{Results}
\begin{figure}
\begin{center}
\begin{tabular}{cc}
    \begin{overpic}[scale=.105]{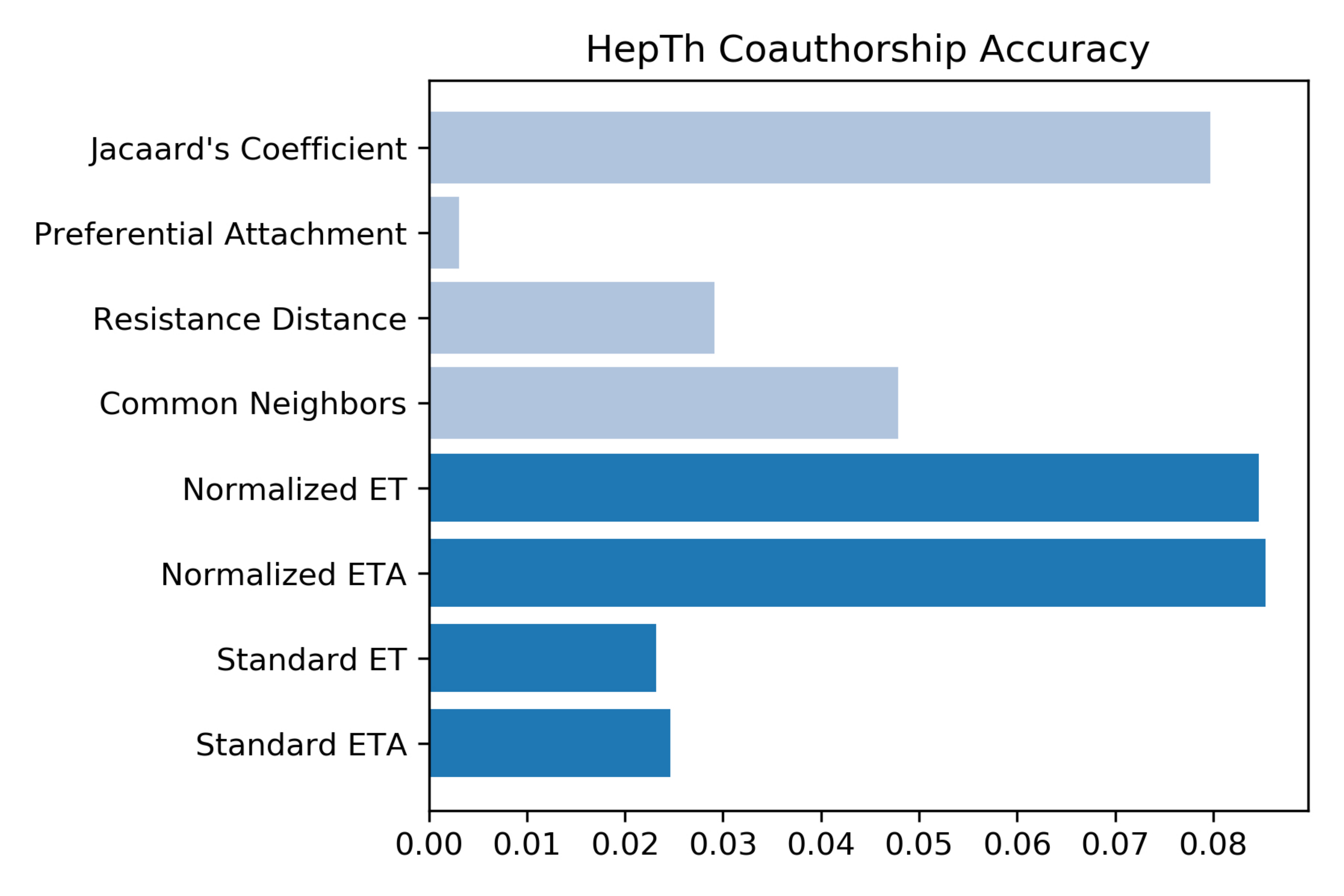}
    \end{overpic} &
    \begin{overpic}[scale=.105]{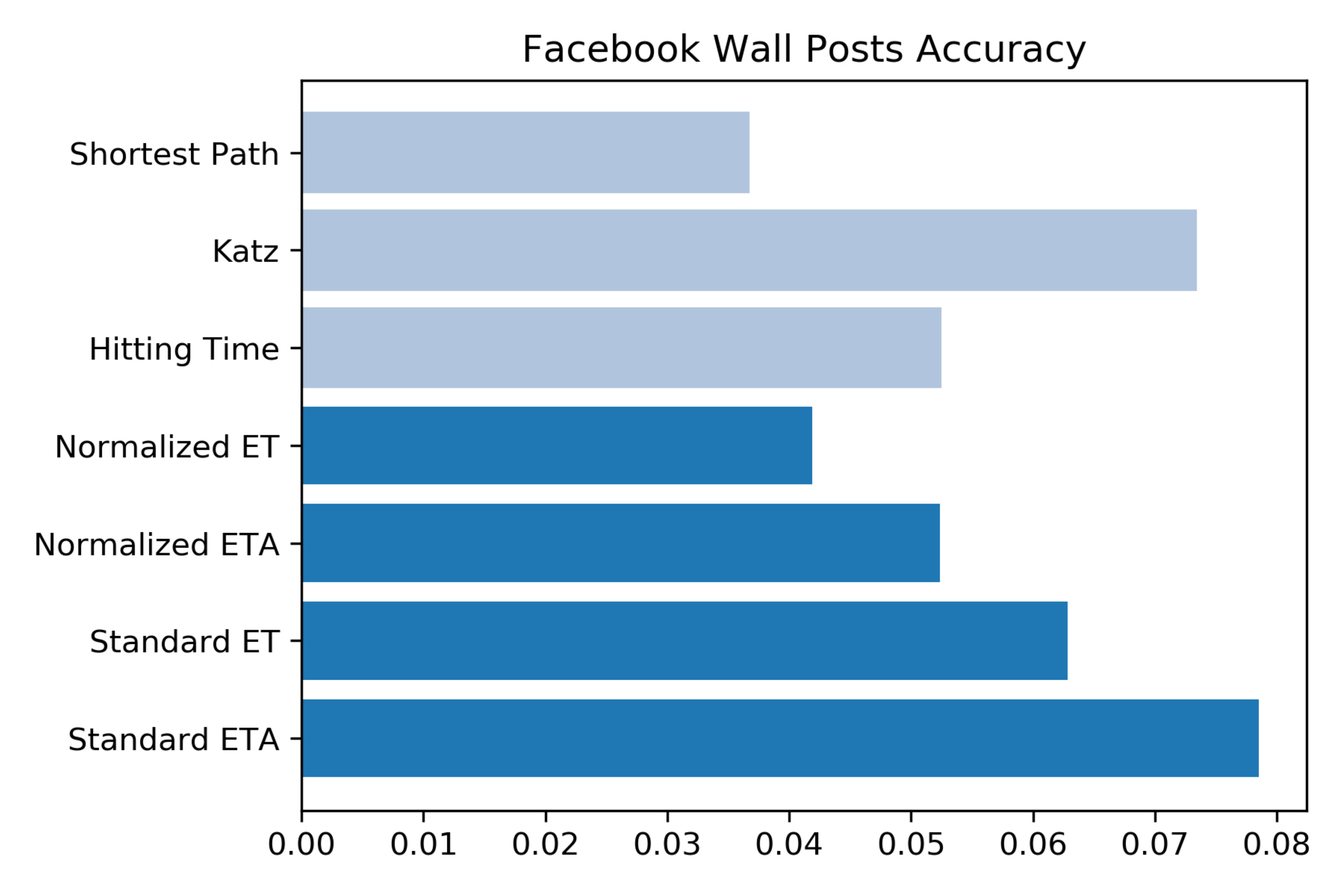}
    \end{overpic}
\end{tabular}
\caption{Left: The undirected HepTh coauthorship network, consisting of authors of high-energy physics papers, is analyzed using the standard methods in Section \ref{sec:undirected} (light blue) and the Effective Transition (ET) methods (dark blue). The Normalized and Standard Effective Transition Approximations (ETA) methods (dark blue) are described in Section \ref{sec:approx}. Right: The directed Facebook Wall Post network consisting of messages between Facebook users is analyzed using the standard methods described in Section \ref{sec:directed} and the Effective Transition methods. Both networks are smaller versions of the larger networks considered later in Figure \ref{Tab:1}.}\label{Fig:ET1}
\end{center}
\end{figure}

Figure \ref{Fig:ET1} shows the results for two networks. On the left are the results of using the predictors from Section \ref{sec:undirected} and the Effective Transition (ET) predictors from Section \ref{sec:ET} on the \emph{small} HepTh coauthorship network \cite{hepth}. This is an undirected network consisting of authors of high-energy physics papers in which coauthors are linked. The network is only a fraction of the size of the \emph{large} HepTh network considered in Figure \ref{Tab:1} (left) as it has $n=387$ nodes and $m=5525$ edges. For this network, the Normalized ET method performs fairly well surpassing all others. The Standard ET method, however, is among the lowest only surpassing Preferential Attachment in accuracy. (The predictors Normalized ETA and Standard ETA in Figure \ref{Fig:ET1} are described in the following section.)

Figure \ref{Fig:ET1} (right) shows the results for the \emph{small} Facebook Wall Posts network, which is a directed network representing messages between Facebook users \cite{fb}. Here the effective transition methods of Section \ref{sec:ET} are compared against the standard prediction methods from Section \ref{sec:directed}. This network is a subset of the full Facebook Wall Post network considered in Figure \ref{Tab:1} (right) having $n=303$ nodes and $m=761$ edges. On this network the Standard ET method is fairly competitive being outdone only by the Katz method of all the standard predictors described in Section \ref{sec:directed}. The Normalized ET method is only better, in this case, than the Shortest Path predictor.

The smaller networks were created for both the HepTh coauthorship network and the Facebook Wall Post network by taking the largest (strongly) connected component of the first 1000 nodes to form in each network, respectively.

As can be seen, for some networks the normalized effective transition predictor achieves a higher accuracy while in others the standard effective transition predictor is more accurate. It is unknown as to when one ET predictor will be more accurate and why.

\section{Approximating Effective Transitions}\label{sec:approx}

Although our algorithm is competitive in accuracy, it is temporally expensive due to the inverse in the computation of each isoradial reduction (cf. Equation \ref{eqn:IRR}). For a graph with transition matrix $M\in\mathbb{R}^{n\times n}$, the temporal complexity of computing $\mathcal{E}(M)$ using Equation \eqref{eqn:def} is $\mathcal{O}(n^5)$. This is because we must invert ${n\choose 2}=n(n-1)/2\in\mathcal{O}(n^2)$ matrices of size $(n-2) \times (n-2)$, where each inverse has temporal complexity $\mathcal{O}(n^3)$. As a result, this method cannot be effectively used on very large networks, at least not directly. This is the reason for considering the smaller networks described in Figure \ref{Fig:ET1}.

\subsection{Reducing Temporal Complexity}
To address this issue of high temporal complexity, we introduce an approximation of the effective transition predictor. The idea is instead of computing the asymptotic probability of eventually transitioning from node $i$ to $j$ before returning to $i$ we compute the probability of transitioning from node $i$ to node $j$ before returning to node $i$ in some fixed $\ell>0$ steps.

Since the nodes that lie only on walks of length greater than $\ell$ between vertices $i$ and $j$ do not contribute to this $\ell$-step approximation, these nodes can be safely ignored in our approximation. To determine which nodes do and do not contribute to this approximation we use a breadth-first search to construct a distance matrix $\Delta=[\delta_{ij}]\in\mathbb{N}^{n\times n}$, where $\delta_{ij}$ is the distance from node $i$ to $j$ in the network.

Here, node $k$ is unnecessary in our $\ell$-step approximation of the eventual transition probability from node $i$ to $j$ if $\delta_{ik}+\delta_{kj}>\ell$. In this case we remove node $k$ from the network when computing this probability. To make this precise we define the set
\[
\Gamma_{ij}^\ell=\{k\in N:\delta_{ik}+\delta_{kj}\leq\ell,\delta_{jk}+\delta_{ki}\leq\ell\}\cup\{i,j\} \ \ \text{for} \ \ i,j\in N \ \text{with} \ \ i\neq j,
\]
which are all vertices on walks of length less than or equal to $\ell$ from node $i$ to $j$ or from node $j$ to $i$ together with both $i$ and $j$. This allows us to define the following approximation of our effective transition method.

\begin{definition} \textbf{($\mathbf{\ell}$-Step Approximations)} For the transition matrix $M$ of $G=(N,E)$, the set $S=\{i,j\}$, and the positive integer $\ell<\infty$, let $\tilde{S}$ be the complement of $S$ in $\Gamma_{ij}^\ell\subseteq N$. Then
\begin{equation}
\label{eq:lstep}
\mathcal{I}_S^\ell(M)=M_{SS}+\rho(M)^{-1}M_{S\tilde{S}}\left(\sum\limits_{k=0}^\ell\left(\rho(M)^{-1} M_{\tilde{S}\tilde{S}}\right)^k\right)M_{\tilde{S}S}
\end{equation}
is the \emph{$\ell$-step approximation} of $\mathcal{I}_S(M)$. The matrix $\mathcal{E}^\ell(M)=[\varepsilon_{ij}^\ell]$ given by
\begin{equation}
\varepsilon_{ij}^\ell=
\begin{cases}
    \mathcal{I}^\ell_{\{i,j\}}(M)_{12} &\text{if} \ i< j\\
    \mathcal{I}^\ell_{\{i,j\}}(M)_{21} &\text{if} \ i> j\\
\sum\limits_{k>i} \mathcal{I}^\ell_{\{i,k\}}(M)_{11}+\sum\limits_{k<i} \mathcal{I}^\ell_{\{i,k\}}(M)_{22} &\text{if} \ i=j
\end{cases}
\end{equation}
is the $\ell$-step approximation of the effective transition matrix $\mathcal{E}(M)$.
\end{definition}

\begin{figure}
\begin{center}
\begin{tabular}{cc}
    \begin{overpic}[scale=.105]{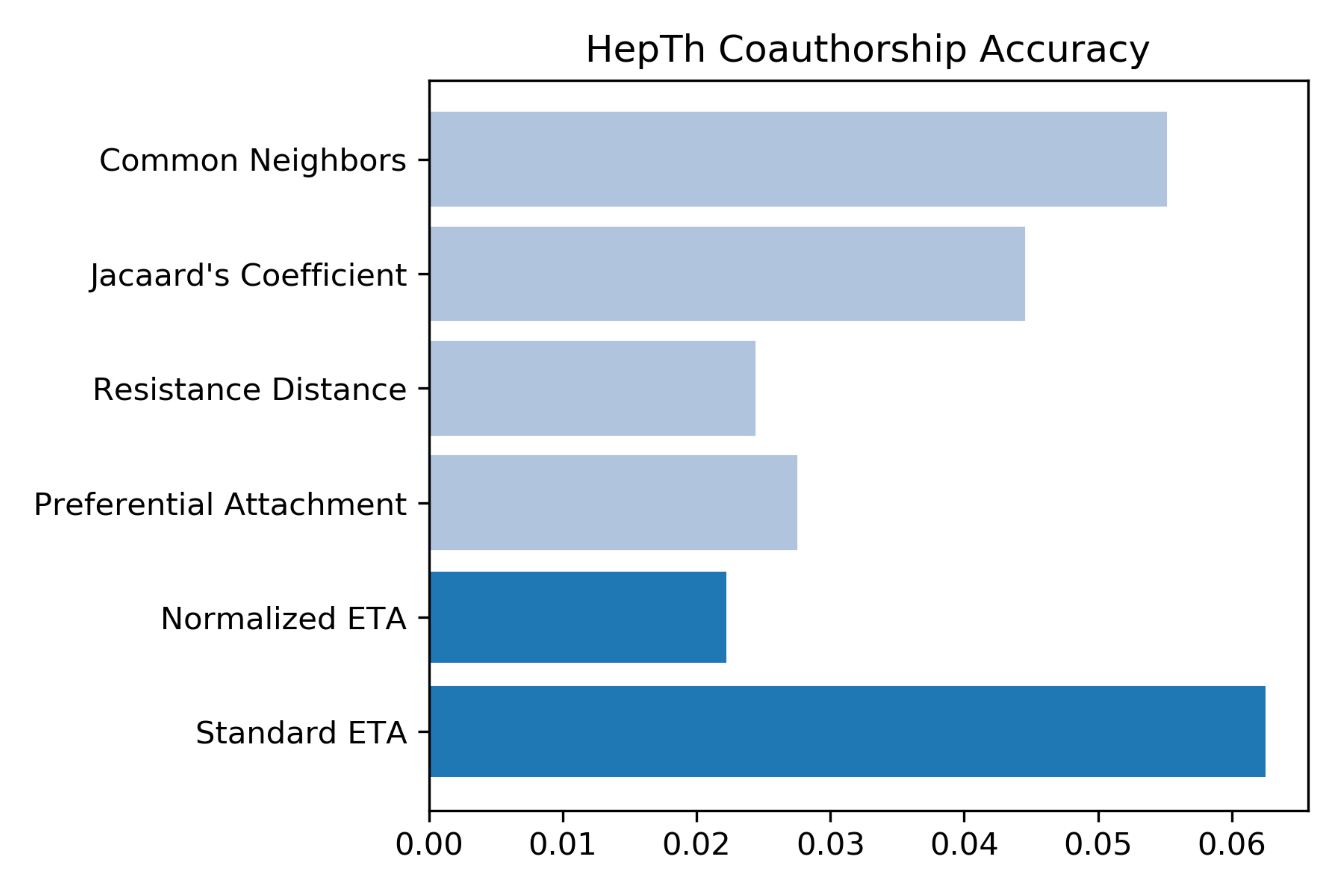}
    \end{overpic} &
    \begin{overpic}[scale=.105]{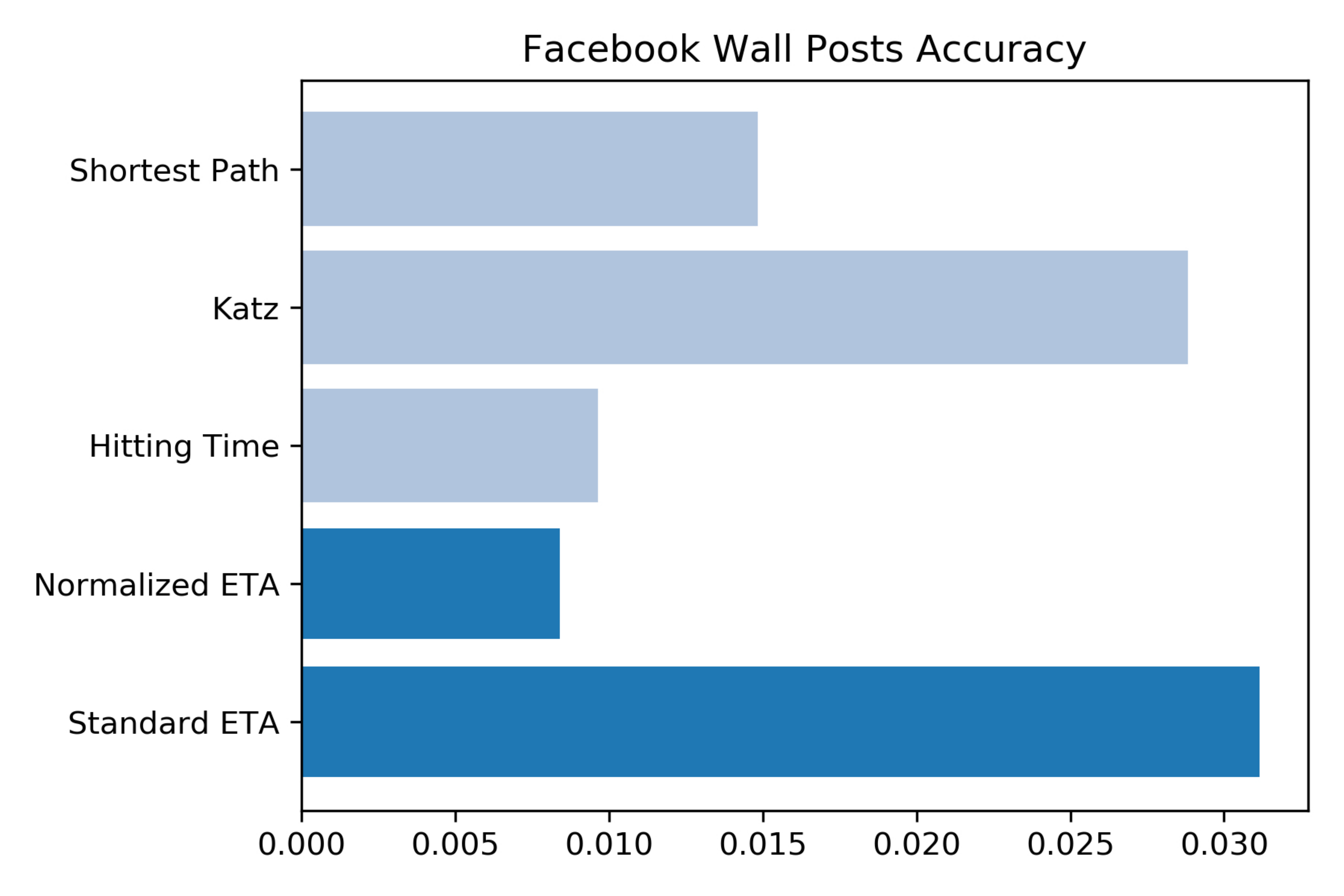}
    \end{overpic}
\end{tabular}
\caption{Left: The full HepTh coauthorship network, consisting of authors of high-energy physics papers, is analyzed using the methods in Section \ref{sec:undirected} (light blue) and the Effective Transition Approximation methods (dark blue) using $\ell=3$. Right: The full Facebook Wall Post network consisting of messages between Facebook users is analyzed using the methods of Section \ref{sec:directed} and the Effective Transition Approximation methods (dark blue) using $\ell=3$.}\label{Tab:1}
\end{center}
\end{figure}

This $\ell$-step approximation allows us to define the following link predictor.

\vskip .2cm
\emph{Approximate Effective Transition:} Suppose $G = (N,E)$ is a (strongly) connected graph with transition matrix $M$. For $\ell>0$ the $\ell$-step approximate effective transition score is given by
\begin{equation}
score(i,j) = \varepsilon^\ell_{ij}.
\end{equation}
\vskip .2cm

The following theorem justifies the notion that $\mathcal{E}^\ell(M)$ is an approximation of $\mathcal{E}^\ell(M)$ and describes a few of its properties. Its proof is given in the Appendix. For two matrices $A=[a_{ij}]$ and $B=[b_{ij}]$ of the same size we write $A\leq B$ if $a_{ij}\leq b_{ij}$ for all entries of $A$ and $B$.

\begin{theorem}\label{thm:lstep} \textbf{(Properties of $\ell$-Step Approximations)}
If $M$ is a nonnegative irreducible matrix and $\ell>0$ then\\
(i) $\lim_{\ell \rightarrow \infty} \mathcal{E}^\ell(M)=\mathcal{E}(M)$;\\
(ii) the sequence $\{\mathcal{E}^\ell(M)\}_{\ell\geq0}$ is nondecreasing, i.e. $\mathcal{E}^\ell(M) \leq \mathcal{E}^{\ell+1}(M)$; and\\
(iii) the temporal complexity of computing $\mathcal{E}^\ell(M)$ is $\mathcal{O}(n^{2.373}+s^{2.373} n)$, where\\ $s=\frac{1}{n}\sum_{i=1}^n |\Gamma_\ell(i)|$ and $\Gamma_\ell(i)$ is the set of all nodes within $\ell$ steps of node $i$.
\end{theorem}

The idea behind the $\ell$-step approximation is that long walks between nodes contribute comparitively little to an effective transition score. Instead of computing the full effective transition score, i.e. the score when $\ell=\infty$, we compute the score attained by transitioning in a finite number of steps. For a better approximation of the true effective transition score we simply increase the number of steps. Importantly, for small $\ell$ the size of the matrices in Equation \eqref{eq:lstep} are also small and the sum in this equation contains only $\ell+1$ terms leading to a lower temporal complexity This allows us to compute an approximate effective transition (ETA) score for much larger networks. (See Section \ref{sec:approxres}.)

\begin{figure}
\begin{center}
\begin{tabular}{cc}
    \begin{overpic}[scale=.105]{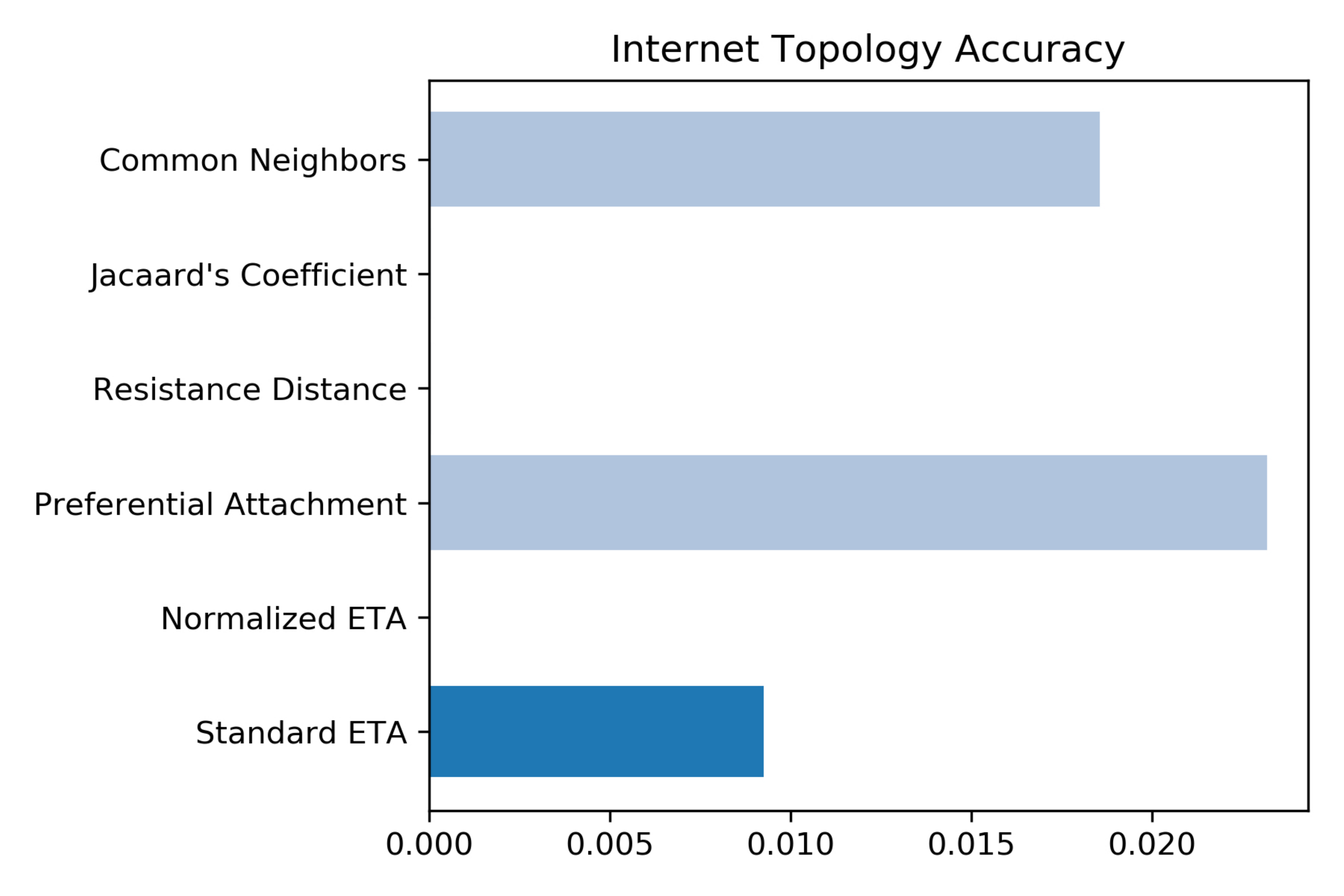}
    \end{overpic} &
    \begin{overpic}[scale=.105]{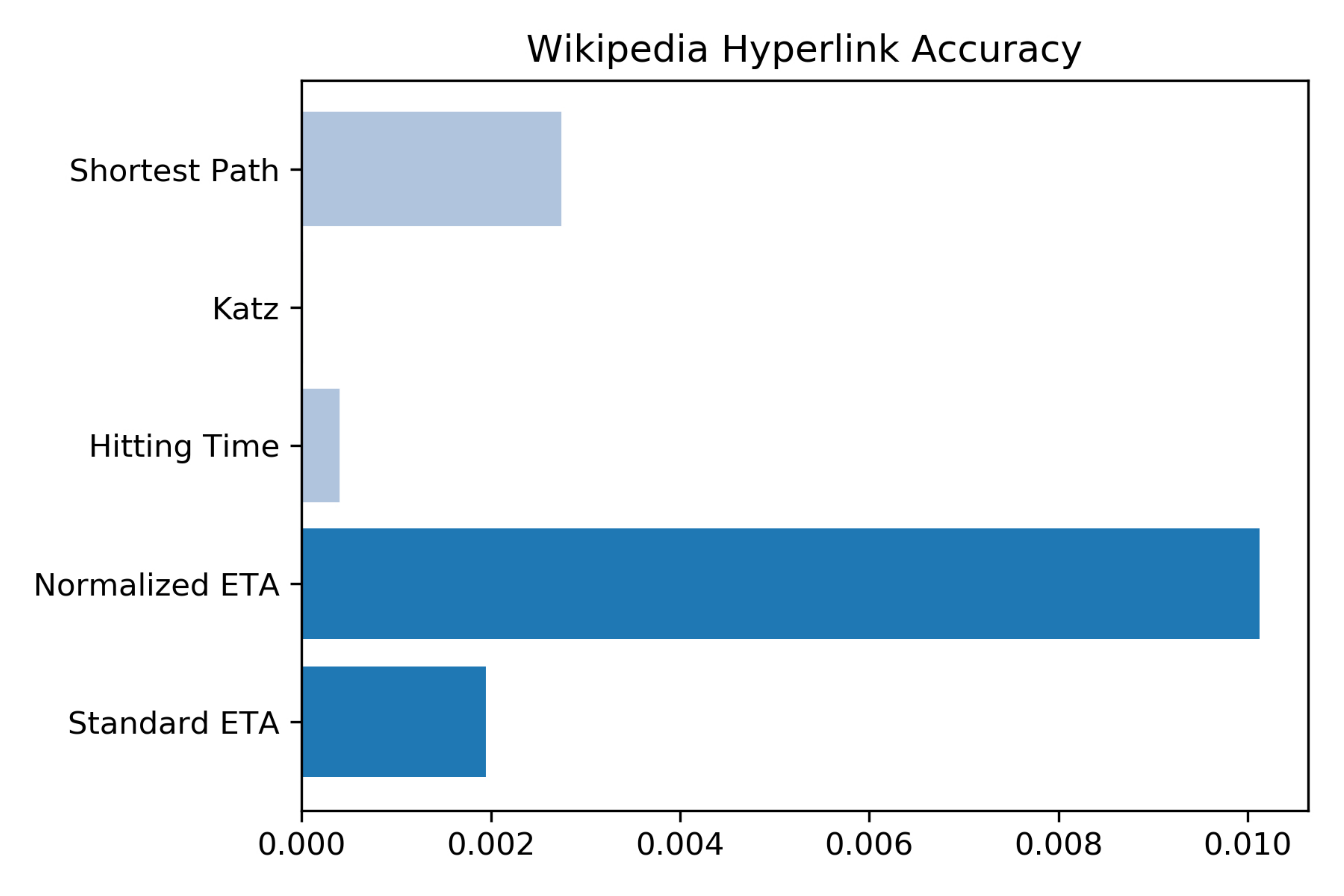}
    \end{overpic}
\end{tabular}
\caption{Left: The undirected Internet Topology network, consisting of computers and the connections between them is analyzed using methods in Section \ref{sec:undirected} (light blue) and the Approximate Effective Transition methods (dark blue) for $\ell=2$. Right: The directed Wikipedia Hyperlink network is analyzed using methods in Section \ref{sec:directed} and the Approximate Effective Transition methods for $\ell=2$.}\label{Fig:Wiki}
\end{center}
\end{figure}

\subsection{Results of Approximate Transition Method}\label{sec:approxres}
Here we describe the results of using the approximate effective transition method on a number of large real-world networks. In Figure \ref{Tab:1} (left) we again consider the now ``full" HepTh network. This network is much larger than the one considered in Figure \ref{Fig:ET1} (left) having $n=2190$ nodes and $m=44590$ edges. For this network it is not feasible to run either the Normalized or Standard ET methods, only their approximations. These are shown as Normalized ETA and Standardized ETA in the figure and are approximated using $\ell=3$ for both. While the Normalized ETA is the least accurate of the predictors the Standard ETA method does better than any other. One might assume that the same type of results would be found in Figure \ref{Fig:ET1} (left) as this is a smaller version of the network. However, this is not the case as the Normalized and Standard approximation using $\ell=3$ for both on the smaller HepTh network are both small but not the least predictive of the methods. This suggests that the way new links form in the smaller HepTh network is different from the way links form in the full HepTh network.

In Figure \ref{Tab:1} (right) we examine the larger Facebook Wall Post network consisting of $n=295$ nodes and $m=10011$ edges (c.f. Figure \ref{Fig:ET1}, right). Here the Normalized ETA and Standard ETA methods are again done using $\ell=3$ and $\ell=3$, respectively. As in Figure \ref{Tab:1} (left) the Standard ETA outperforms all other methods while the Normalized ETA has the least accuracy. What is perhaps surprising is the fact that this is not the case in the smaller version of this network where the Normalized and Standard ETA methods outperform all other methods, with the exception that the Katz method out performs the Normalized ETA method.

It is worth noting that both networks in Figures \ref{Fig:ET1} and \ref{Tab:1} are social networks. For the sake of diversity we also consider the Internet Topology network \cite{top} and the Wikipedia Hyperlink network \cite{wiki} shown in Figure \ref{Fig:Wiki}, which are technological and information networks, respectively. As these network are relatively large with $n=2864$ and $n=2809$ nodes, respectively, we use only the approximate effective transition predictors on them. In Figure \ref{Fig:Wiki} the predictors that outperform the other predictors are quite different and much less accurate than those in Figures \ref{Fig:ET1} and \ref{Tab:1} suggesting that these networks experience a different kind of growth than the two social networks.

\begin{figure}
\begin{center}
    \begin{overpic}[scale=.15]{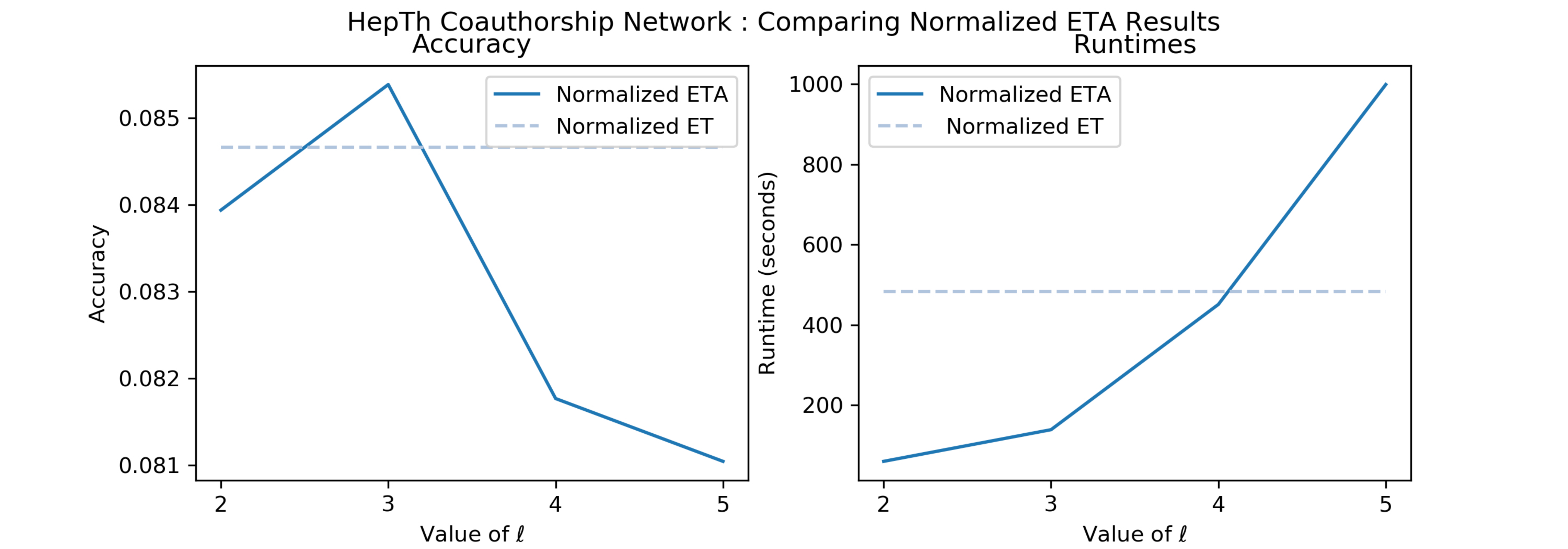}
    \end{overpic}
    \caption{Left: The accuracy of the approximate normalized effective transition method on the Hep-Th network considered in Figure \ref{Fig:ET1} (left) for $\ell=2,\dots,5$ is shown. The solid blue and dashed blue lines indicate the approximate and unapproximated accuracy using this method on the smaller Hep-Th network, respectively. Right: The time to compute each approximate and unapproximated effective transition score for $\ell=2,\dots,5$ is shown.}\label{Fig:Approx}
\end{center}
\end{figure}

We note that the accuracy of the effective transition approximation depends, in the networks we have considered, on $\ell$ but in a surprising way. As shown in Figure \ref{Fig:Approx} it is usually optimal to choose $\ell$ quite small, in which case our approximation has significantly lower temporal complexity. Choosing the optimal $\ell$ also tells us something about how quantities such as influence, information, contagions, energy, etc. may be passed through the network. Specifically, the $\ell$ that maximizes the accuracy of an effective transition predictor gives us an idea for the optimal neighborhood size at which connections are made, at least on average, within the network.

\section{Weighted Graphs}\label{sec:weight}
The edges of a network can be either weighted or unweighted. Up to this point, all graphs we have considered have been assumed to have unweighted edges. In the case that a graph $G$ has weighted edges, we refer to it as a \emph{weighted graph} and write $G=(N,E,\omega)$, where $\omega : E \rightarrow \mathbb{R}$ is a function that assigns a nonzero weight to each edge. The weights of the edges given by $\omega$ typically measure the \emph{strength} of these interactions. For example, in social networks weights can correspond to the frequency of interaction between individuals. In food web networks weights measure energy flow, and in traffic networks weights measure how often specific routes are used (see \cite{Newman10} for examples and applications).

For a weighted graph $G=(N,E,\omega)$ there is a corresponding \emph{weighted adjacency matrix} $W=[w_{ij}]\in\mathbb{R}^{n\times n}$ with entries
\[
w _{ij} = \left\{
\begin{array}{ll}
\omega(e_{ij})\neq 0 & \text{if  }e_{ij}\in E \\
0 & \text{otherwise}.
\end{array}
\right.
\]

Link prediction on weighted networks differs from link prediction on unweighted networks because few if any predictors are designed to incorporate edge weights into the score functions (see, for instance, \cite{link_prediction_social_networks}). The versatility of the effective transition predictor, however, allows for edge weights to affect the link-prediction scores. That is, if $G$ is (strongly) connected and the matrix $W$ is nonnegative we can compute the effective transition matrix $\mathcal{E}(W)=[\varepsilon_{ij}]$ or its $\ell$-step approximation $\mathcal{E}^\ell(W)=[\varepsilon^\ell_{ij}]$ to score the edges and potential edges of $G$.

\begin{figure}
\begin{center}
    \begin{overpic}[scale=.1075]{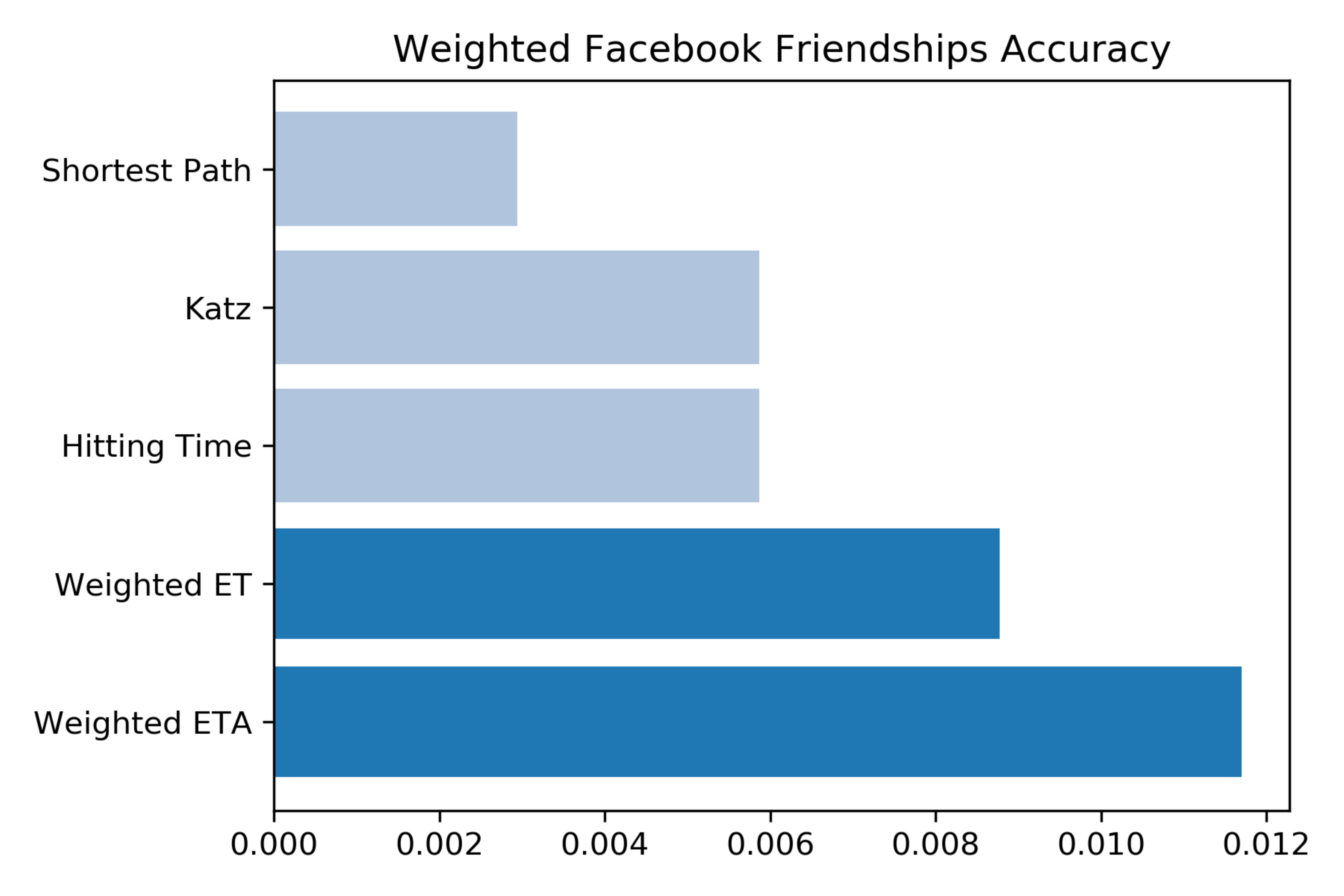}
    \end{overpic}
\caption{A weighted Facebook friendship network, consisting of edges weighted by how frequently users post on eachother's walls, is analyzed using the network's weighted adjacency matrix and the effective transition method. The results for the standard method from Section \ref{sec:undirected} applied to the unweighted version of the graph are shown in light blue. The results using the ``weighted" effective transition method and its approximation with $\ell=2$ are shown in dark blue.}\label{Fig:Weighted}
\end{center}
\end{figure}

Figure \ref{Fig:Weighted} shows the results for a weighted network of Facebook friendships with $n=519$ nodes and $m=1389$ edges. Here weights correspond to how frequently a user posts on another user's wall. Hence, the network is a directed weighted network. Both the weighted ET method and its approximation with $\ell=2$ outperform the other standard methods that are used on the unweighted version of the network. In fact, these ET methods significantly outperform these other methods suggesting that, at least for this network, weights have a significant impact on the network's growth.

\section{Conclusion}\label{sec:conc}

In this paper we present a link prediction framework based on the notion of effective transitions that can be used on most any kind of network including directed or undirected and weighted or unweighted networks. The fact that this method works on weighted networks is quite novel as very few if any link predictors are designed to take into account the weights of network interactions.

Because of the high temporal cost of this method we also devise an $\ell$-step approximation of this method that scales to much larger networks and, in many cases, out performs the original effective transition method. Currently it is an open question as to what $\ell$-step approximation gives the best accuracy for predicting links in a given network and what this says about the growth of the network.

It is also worth mentioning that our standing assumption in this paper is that a network under consideration be (strongly) connected or equivalently that its associated transition matrix is irreducible. In fact, effective transition ``probabilities" are still defined if this condition does not hold. However, in certain situations the isoradial reductions used to compute the probabilities can fail to exist and other considerations must be taken to compute the associated effective transition matrix. Even if the network is not strongly connected the $\ell$-step approximation still exists and can be computed efficiently, at least for small $\ell$.

\section*{Appendix}

Here we prove Theorem \ref{thm:eql}, Proposition \ref{prop:stoch}, and Theorem \ref{thm:lstep}. We begin with Theorem \ref{thm:eql}, which requires the following corollary, which is an immediate consequence of Theorem 1.3 in \cite{BWBook}.

\begin{corollary}\label{cor:unique}
\textbf{(Uniqueness of Sequential Reductions)} Let $G=(N,E)$ be a (strongly) connected graph with transition matrix $P$. If $N\supset N_1\supset\cdots\supset N_m$ then the sequence of isoradial reductions
\[
\mathcal{I}_{N_m}(\mathcal{I}_{N_{m-1}}(\cdots\mathcal{I}_{N_1}(P)\cdots))=\mathcal{I}_{N_m}(P).
\]
\end{corollary}

That is, in a sequence of isoradial reductions the result only depends on the final set $N_m$ over which the matrix is reduced. To prove Theorem \ref{thm:eql} we use the fact that the isoradial reduction $\mathcal{I}_{\{i,j\}}(P)$ over two vertices can be achieved by a sequence of $n-2$ reductions each of which removes a single row and column from the matrix.

\begin{proof}
Suppose the row stochastic matrix $P=[p_{ij}]\in\mathbb{R}^{n\times n}$ is irreducible, i.e. is the transition matrix of some (strongly) connected graph $G=(N,E)$. We note that $p_{ij}\geq0$ is the probability of immediately transitioning from node $i$ to node $j$ in $G$. Equivalently, since $P=\mathcal{I}_N(P)$ where $N=\{1,2,\dots,n\}$ then $p_{ij}$ is the probability of eventually transitioning from node $i$ to node $j$ in $G$ before transitioning to any other node $k\in N-\{j\}$. We assume then, by way of induction, that for the set $S=\{1,2,\dots,m\}\subset N$ with $m>2$ and $Q=[q_{ij}]\in\mathbb{R}^{m\times m}$ where $Q=\mathcal{I}_S(P)$ that $q_{ij}\geq 0$ is the probability of eventually transitioning from node $i$ to node $j$ in $G$ before transitioning to any other node $k\in S-\{j\}$.

Letting $S\supset T=\{1,2,\dots,m-1\}$ and $R=[r_{ij}]\in\mathbb{R}^{(m-1)\times(m-1)}$ where $R=\mathcal{I}_{T}(P)$ note that
\[
r_{ij}=q_{ij}+\frac{q_{im}q_{mj}}{1-q_{mm}}
\]
using Equation \eqref{eqn:def}. Furthermore, $q_{ij}$ is the probability of eventually transitioning from node $i$ to node $j$ in $G$ before transitioning to any other node $k\in S-\{j\}=\{1,3,4,5,\dots,m\}$. Also,
\[
q_{im}q_{mj}+q_{im}q_{mm}q_{mj}+q_{im}q_{mm}^2q_{mj}+\cdots=q_{im}q_{mj}\sum_{k=0}^\infty q_{mm}^k=\frac{q_{im}q_{mj}}{1-q_{mm}}
\]
since $q_{mm}<1$, given that $G$ is irreducible, which is the probability of eventually transitioning from node $i$ to node $j$ in $G$ before transitioning to any other node $k\in T-\{j\}=\{1,3,4,5,\dots,m-1\}$. Hence, $r_{ij}$ is the probability of eventually transitioning from node $i$ to node $j$ in $G$ before transitioning to any other node $k\in T-\{j\}$.

By induction we can repeatedly remove a single row and column from $P$ via a sequence of isoradial reductions until we have the $2\times 2$ matrix $\mathcal{I}_{\{1,2\}}(P)$ where $\mathcal{I}_{\{1,2\}}(P)_{12}$ is the probability of eventually transitioning from node $1$ to node $2$ in $G$ before transitioning to node $1$, the only element of the set $\{1,2\}-\{2\}$. Here we are using the fact that the set $N\supset\dots S\supset T\supset\dots\supset \{1,2\}$ so that via Corollary \ref{cor:unique} we always arrive at the unique matrix $\mathcal{I}_{\{1,2\}}(P)$ under this sequence of reductions.

As this argument holds not only for nodes $1$ and $2$ but any pair of indices $i,j\in N$ with $i\neq j$, the first and second parts of Equation \eqref{eqn:stoch} also hold. The theorem then follows from the fact that $1-\mathcal{I}_{\{i,k\}}(P)_{12}=\mathcal{I}_{\{i,k\}}(P)_{11}$ for $i<k$ is the probability of eventually transitioning from node $i$ back to node $i$ without first transitioning to node $k$ and the analogous formula for the case when $i>k$.
\end{proof}

The following proof of Proposition \ref{prop:stoch} relies on the following result of \cite{Smith19}.

\begin{theorem} \label{thm:irr}  \textbf{(Properties of Isoradial Reductions)}
Let $M\in \mathbb{R}^{n\times n}$ be a nonnegative irreducible matrix and let $S\subset N$ in which $|S|=m$. Then
\\
(a) the isoradial reduction $\mathcal{I}_S(M)\in \mathbb{R}^{m\times m}$ exists and is a nonnegative and irreducible matrix with the same spectral radius, i.e. $\rho (M)=\rho(\mathcal{I}_S(M))$; and \\
(b) if $\mathbf{v}\in \mathbb{R}^{n\times 1}$ is the leading eigenvector of $M$ then its projection $\mathbf{v}_S\in \mathbb{R}^{m\times 1}$ to its components indexed by $S$ is the leading eigenvector of $\mathcal{I}_S(M)$.
\end{theorem}

Property (a) in Theorem \ref{thm:irr} states that the spectral radius of a matrix is unaffected by an isoradial reduction. Property (b) states that the leading eigenvector of an isospectral reduction over the set $S$ is simply the projection of the leading eigenvector of the original matrix onto $S$ (see also Theorem 2.2 and 3.1 in \cite{Mey89}, respectively). Using Theorem \ref{thm:irr} we give the following proof of Proposition \ref{prop:stoch}.

\begin{proof} Part (a) of Theorem \ref{thm:irr} together with the proof of Theorem \ref{thm:eql} imply that each isoradial reduction
\begin{equation}\label{eq:red}
\mathcal{I}_{\{i,j\}}(M)=
\left[
\begin{array}{cc}
*&\varepsilon_{ij}\\ [1pt]
\varepsilon_{ji} & *
\end{array}
\right]
\in\mathbb{R}^{2\times 2},
\end{equation}
is an irreducible row stochastic matrix for $i<j$. Hence, both $0<\varepsilon_{ij}$, $\varepsilon_{ji}\leq 1$. Equation \eqref{eqn:stoch} of Theorem \ref{thm:eql} then implies that each non-diagonal entry of $\mathcal{E}(M)$ is strictly positive and $\mathcal{E}(M)$ is therefore nonnegative and irreducible.

To prove parts (ii) and (iii) let $E^{ij}\in\mathbb{R}^{n\times n}$ be the matrix with
\[
E^{ij}_{ii}=1-\varepsilon_{ij}, \ E^{ij}_{ij}=\varepsilon_{ij}, \ E^{ij}_{ji}=\varepsilon_{ji}, \ \text{and} \ E^{ij}_{jj}=1-\varepsilon_{ji}
\]
and all other entries equal to zero for $i<j$. Thus, $\mathcal{E}(M)=\sum_{i<j}E^{ij}$. Similarly, for the leading eigenvector $\mathbf{v}=(v_1,\dots,v_n)$ of $M$ let $\mathbf{v}^{ij}\in\mathbb{R}^{n}$ be the vector with $v^{ij}_i=v_i$ and $v^{ij}_j=v_j$ in which all other entries equal to zero for $i<j$. Hence, $\sum_{i<j}\mathbf{v}^{ij}=(n-1)\mathbf{v}$. By part (b) of Theorem \ref{thm:irr} we have $E^{ij}\mathbf{v}^{ij}=\rho(M)\mathbf{v}^{ij}$ where $\rho(M)>0$ as $M$ is nonnegative and irreducible. Thus,
\begin{equation*}
\mathcal{E}(M)\mathbf{v}=(\sum_{i<j}E^{ij})\mathbf{v}=\sum_{i<j}(E^{ij}\mathbf{v}^{ij})=\sum_{i<j}(\rho(M)\mathbf{v}^{ij})=(n-1)\rho(M)\mathbf{v}.
\end{equation*}
Since $M$ and $\mathcal{E}(M)$ are nonnegative and irreducible the Perron-Frobenius theorem implies that the leading eigenvectors of both matrices are the matrices' unique eigenvectors with strictly positive entries, respectively. Hence, the vector $\mathbf{v}>0$ implying $(n-1)\rho(M)\mathbf{v}$ so that this vector must also be the leading eigenvector of $\mathcal{E}(M)$ and $\rho(\mathcal{E}(M))=(n-1)\rho(M)$.

To prove part (iii) we note that as $P^T$ is also irreducible then by the proof of part (ii) there is a unique leading eigenvector $\mathbf{p}\in\mathbb{R}^n$ of both $P^T$ and $\mathcal{S}(P^T)$ corresponding to the spectral radius $\rho(P)=\rho(\mathcal{S}(P))=1$ with $\sum_{i=1}^n p_i=1$. Since $\mathcal{E}(P^T)=\mathcal{E}(P)^T$ using Equation \ref{eq:red} then $\mathcal{S}(P^T)=\mathcal{S}(P)^T$ implying $P^T\mathbf{p}=\mathbf{p}$ and $\mathcal{S}(P)^T\mathbf{p}=\mathbf{p}$. Taking the transpose of both equations yields $\pi=\pi P$ and $\pi=\pi \mathcal{S}(P)$ where $\pi=\mathbf{p}^T$ and $\sum_{i=1}^n \pi_i=1$. Since $\mathbf{p}$ is unique, the Markov chains associated with $P$ and $\mathcal{S}(P)$ have the same unique stationary distribution.

Also, as $\mathcal{E}(P)=[\varepsilon_{ij}]$ has strictly positive off-diagonal entries then each
\[
(\mathcal{E}(P)^2)_{ij}=\sum_{k=1}^n\varepsilon_{ik}\varepsilon_{kj} \ \text{for} \ i,j\in N=\{1,2,\dots,n\}
\]
is strictly positive. Hence, $\mathcal{E}(P)$ is primitive implying $\mathcal{S}(M)=\frac{1}{n-1}\mathcal{E}(M)$ is also primitive. This completes the proof.
\end{proof}

We now give a proof of Theorem \ref{thm:lstep}.

\begin{proof}
Suppose that $M\in\mathbb{R}^{n\times n}$ is a transition matrix of a (strongly) connected graph $G$. To prove part (i) note that $\rho(M)>0$ and for any subset $S=\{i,j\}\subset N$ the spectral radius of the submatrix $\rho(M_{\bar{S}\bar{S}})<\rho(M)$ (see Chapter 8 \cite{HJ90}). Hence, the matrix $I-\rho(M)^{-1}M_{\bar{S}\bar{S}}$ is invertible with inverse
\[
\left(I-\rho(M)^{-1}M_{\bar{S}\bar{S}}\right)^{-1}=\sum_{k=1}^\infty \left(\rho(M)^{-1}M_{\bar{S}\bar{S}}\right)^k,
\]
since the spectral radius of the matrix $\rho(M)^{-1}M_{\bar{S}\bar{S}}$ is less than 1. Note that the set $\tilde{S}=\tilde{S}(\ell)$, given in Definition \ref{eq:lstep}, is a function of $\ell$. However, for $\ell\geq 2\delta(G)$ where $\delta(G)$ is the diameter of $G$ the set $\tilde{S}(\ell)=\bar{S}$. This is because it is possible to go from node $i$ to any other node $k$ and then to node $j$ in at most $2\delta(G)$ steps since the graph $G$ is (strongly) connected implying $\Gamma_{ij}^\ell=N=\{1,2,\dots,n\}$. Therefore, for $\ell\geq \delta(G)$ we have
\[
\mathcal{I}_S^\ell(M)=M_{SS}+\rho(M)^{-1}M_{S\bar{S}}\left(\sum\limits_{k=0}^\ell\left(\rho(M)^{-1}M_{\bar{S}\bar{S}}\right)^k\right)M_{\bar{S}S}
\]
implying
\begin{align*}
\lim_{\ell\rightarrow \infty}\mathcal{I}_S^\ell(M)=&\lim_{\ell\rightarrow \infty}\left[M_{SS}+\rho(M)^{-1}M_{S\bar{S}}\left(\sum\limits_{k=0}^\ell\left(\rho(M)^{-1}M_{\bar{S}\bar{S}}\right)^k\right)M_{\bar{S}S}\right]\\
&=M_{SS}+\rho(M)^{-1}M_{S\bar{S}}\left(\sum_{k=0}^\infty\left(\rho(M)^{-1}M_{\bar{S}\bar{S}}\right)^k\right)M_{\bar{S}S}\\
&=M_{SS}+\rho(M)^{-1}M_{S\bar{S}}\left(I-\rho(M)^{-1}M_{\bar{S}\bar{S}}\right)^{-1}M_{\bar{S}S}\\
&=M_{SS}+M_{S\bar{S}}\left(\rho(M) I-M_{\bar{S}\bar{S}}\right)^{-1}M_{\bar{S}S}\\
&=M_{SS}-M_{S\bar{S}}\left(M_{\bar{S}\bar{S}}-\rho(M) I\right)^{-1}M_{\bar{S}S}\\
&=\mathcal{I}_S(M)
\end{align*}
completing the proof of part (i).
To prove part $(ii)$ for $S=\{i,j\}$ let $\ell\geq 1$ and let $P,Q\in\mathbb{R}^{n\times n}$ where
\[P_{ab}=
\begin{cases}
M_{ab} &\text{if } a,b\in\Lambda_{ij}^\ell \\
0 &\text{otherwise}
\end{cases}
\ \
and
\ \
Q_{ab}=
\begin{cases}
M_{ab} &\text{if } a,b\in\Lambda_{ij}^{\ell+1} \\
0 &\text{otherwise}.
\end{cases}
\]
Then $0\leq P\leq Q$ since $\Lambda_{ij}^\ell\subseteq \Lambda_{ij}^{\ell+1}$ implying $P_{SS}=Q_{SS}$, $0 \leq P_{S\bar{S}} \leq Q_{S\bar{S}}$, $0 \leq P_{\bar{S}\bar{S}} \leq Q_{\bar{S}\bar{S}}$, and $0 \leq P_{\bar{S}S} \leq Q_{\bar{S}S}$. Therefore,
\begin{align}
\mathcal{I}^\ell_{S}(M)=&P_{SS}+\rho(M)^{-1}P_{S\bar{S}}\left(\sum_{k=0}^\infty\left(\rho(M)^{-1}P_{\bar{S}\bar{S}}\right)^k\right)P_{\bar{S}S}\leq\\
                        &Q_{SS}+\rho(M)^{-1}Q_{S\bar{S}}\left(\sum_{k=0}^\infty\left(\rho(M)^{-1}Q_{\bar{S}\bar{S}}\right)^k\right)Q_{\bar{S}S}=\mathcal{I}^{\ell+1}_{S}(M).
\end{align}
To prove part (iii) note that this algorithm can naturally be divided into two steps: calculating $\Gamma_\ell (i)$ for each node $i \in N$ and preforming an approximated isoradial reduction over each relevant pair of nodes $i,j \in N$.

To start assume that $\ell \ll n$ implying $\ell^2 \in \mathcal{O}(1)$ and let $B(\ell) = \sum_{k=0}^\ell M^k.$
By definition, for nodes $i,j \in N$ if $j \in  \Gamma_\ell (i)$ there exists $ k \in \mathbb{N}$ such that $k \leq \ell$ and $M^k_{ij} > 0$. Thus  $\Gamma_\ell (i)$ can be expressed as the set of all all $j \in N$ such that $B(\ell)_{ij} > 0$. With this, the temporal complexity of determining $\Gamma_\ell (i)$ for all nodes $i \in N$ reduces to computing $B(\ell)$. This computation requires $\ell^2$ matrix multiplications which has complexity $\mathcal{O}(n^{2.373})$ (see ,for instance, \cite{LeGal14}).

Next we approximate the isoradial reductions in a double for loop, the first loop iterates through each node in the network $i$ and the second loop iterates over  each node $j \in \Gamma_\ell (i)$. The average number of iterations in the second loop is given by $s =\frac{1}{n}\sum_{i=1}^n |\Gamma_\ell(i)|$.
Each step of the double for loop involves computing Equation \eqref{eq:lstep}, where $|S|=2$ and $|\tilde{S}|=s$ resulting in a complexity of $\mathcal{O}(s^{2.373})$.
Combining all steps of the algorithm yields an overall complexity of $\mathcal{O} (n^{2.373}+s^{2.373}n)$ as desired.
\end{proof}

\bibliography{EffTransReferences}{}
\bibliographystyle{siamplain}

\end{document}